\documentclass[aps,showpacs,preprint,epsf,epsfig, nofootinbib,titlepage]{revtex4-1}
\usepackage{graphicx}
\usepackage{dcolumn}
\usepackage{bm}
\usepackage{hyperref}
\usepackage{epsfig,graphicx,color,appendix}
\usepackage{multirow}
\usepackage{array}
\usepackage{tabularx}
\usepackage{capt-of}
\usepackage{amsmath}
\usepackage{caption}
\usepackage[countmax]{subfloat}

\usepackage{footmisc}
\renewcommand{\thefootnote}{\fnsymbol{footnote}}
\usepackage{hhline}
\begin{document}

\def\|{\vert}
\def\ol{\bar}
\def\ov{\bar}
\def\be{\begin{eqnarray}}
\def\en{\end{eqnarray}}
\def\non{\nonumber}
\def\t{\times}
\def\lb{\Lambda}
\def\om{\Omega}
\def\omc{\Omega_c}
\def\oc{\Omega_c^0}
\def\xic{\Xi_c}
\def\xicd{\Xi_c^{'}}
\def\x{\Xi}
\def\si{\Sigma}
\def\k{{K_1}}
\def\k1d{\underline{K}_{1}}
\def\kb{\bar{K}_{1}}
\def\kbd{\underline{\bar K}_{1}}
\def\ra{\rangle}
\def\A{{\cal A}}
\def\B{{\cal B}}
\def\L{{\cal L}}
\def\ben{\begin{eqnarray}}
\def\en{\end{eqnarray}}
\def\non{\nonumber}
\def\la{\langle}
\def\ra{\rangle}
\def\t{\times}
\def\pp{{\prime\prime}}
\def\nc{N_c^{\rm eff}}
\def\vp{\varepsilon}
\def\hep{\hat{\varepsilon}}
\def\up{\uparrow}
\def\dw{\downarrow}
\def\vma{{_{V-A}}}
\def\vpa{{_{V+A}}}
\def\smp{{_{S-P}}}
\def\spp{{_{S+P}}}
\def\J{{J/\psi}}
\def\ov{\bar}
\def\Lqcd{{\Lambda_{\rm QCD}}}
\long\def\symbolfootnote[#1]#2{\begingroup\def\thefootnote{\fnsymbol{footnote}}\footnote[#1]{#2}\endgroup}
\def\lsim{ {\ \lower-1.2pt\vbox{\hbox{\rlap{$<$}\lower5pt\vbox{\hbox{$\sim$}
}}}\ } }
\def\gsim{ {\ \lower-1.2pt\vbox{\hbox{\rlap{$>$}\lower5pt\vbox{\hbox{$\sim$}
}}}\ } }

\font\el=cmbx10 scaled \magstep2{\obeylines\hfill \today}

\vskip 1.5 cm
\title{ \Large Charmed Axial Vector and Pseudoscalar Mesons \\ Emitting Decays of Bottom Meson in NRQM }
 
\author{\bf Neelesh Sharma$^{\dagger}$  and Rohit Dhir$^{\ddagger}$ }
\email { {$^{\dagger}$ nishu.vats@gmail.com}, {$^{\ddagger}$dhir.rohit@gmail.com }}
\medskip
\affiliation{\sl $^{\dagger}$SRM Research Institute, SRM University Chennai-603203,India.\\
$^{\ddagger}$Department of Physics and Nanotechnology,\\ SRM University Chennai-603203, India.\\
\\
}
\bigskip
\bigskip
\begin{abstract}
Two body nonleptonic weak decays of bottom mesons to a pseudoscalar meson and an axial-vector meson involving charmed states are studied using the non-relativistic quark model. We calculate the branching ratios of these decays by employing the factorization hypothesis. Obtained results are in good agreement with the existing experimental data. We also calculated the branching ratios in the light of heavy quark symmetry constraints and compared with the exiting theoretical analyses.
\end{abstract}
\pacs{13.25.Ft, 14.40.Cs, 14.40.Ev}
\maketitle
\thispagestyle{empty}
\setcounter{page}{1}

\section{INTRODUCTION}

The phenomenological analyses of bottom (\textit{B}) meson decays provide exciting opportunities to test several models and approaches within and beyond the standard model (SM). The understanding of \textit{B} physics serve as a tool to study the interplay of strong and weak interaction dynamics in the SM and also as a test of the new physics beyond the SM. The time to time comparison between theoretical predictions and experimental results helps in better understanding of hadronic structure of heavy bound states. The theoretical interpretation of nonleptonic decays of heavy flavor hadrons is still under progress for being non-perturbative in nature. The factorization hypothesis has been successfully used to study such decays as it suits the description of \textit{B} decaying to heavy mesons \cite{1,2,3,4,5,6,7,8}. Also, the mass of \textit{b}-quark is much heavier than QCD scale, where the dynamics becomes much simpler in the light of heavy quark symmetries \cite{3,4,7}. The heavy quark symmetry (HQS) proves to be a very useful tool to provide symmetry relations for heavy meson decays, however, one still have to use some model to obtain the explicit expression for decay rates. There are several models \textit{like} Bauer, Stech \& Wirbel (BSW) \cite{3,4}, Isgur-Scora-Grinstein-Wise (ISGW) \cite{5,6}, and covariant light front (CLF) model \cite{9} which can effectively be used in light of heavy quark effective theory (HQET). Unlike the present study, in the heavy to light decays the final state mesons have large energy, where the approaches like light cone sum rules, soft collinear effective theory (SCET) and perturbative QCD have worked reasonably well to understand the experimental data \cite{7,8}. Several theoretical frameworks based on factorization hypothesis, relativistic quark model, heavy quark effective theory, covariant light front (CLF) approach, perturbative QCD, \textit{etc}. has been employed to study \textit{charmless} axial-vector meson emitting decays of \textit{B} mesons \cite{9,10,11,12,13,14,15,16}. It is worth remarking here that the dominant modes in charmed sector of these decays proceed mainly through the tree level diagrams and thus, are least influenced by penguin pollution. Thus, \textit{charmed} \textit{p}-wave meson emitting decays are considered using factorization scheme in non-relativistic quark models, CLF approach, SCET, perturbative QCD based modeling and HQET \textit{etc.} \cite{16,17,18,19,20,21,22,23,24,25,26,27}.  

On the experimental side, recent observations, especially, of many strange charm resonances and many proposed experiments (see for review \cite{28}) has revived the interests of hadronic physicists to study the orbitally excited mesons. In weak decays sector, many of decay modes involving charm meson in the final state \textit{e.g.} $B^{-} \to D^{0} a_{1}^{-} $, $B^{-} \to \pi ^{-} \underline{D}_{1}^{0} $, $B^{-} \to \pi ^{-} \chi _{c1} $, $\bar{B}^{0} \to D^{+} a_{1}^{-} $, $\bar{B}^{0} \to \chi _{c1} \pi ^{0} $, $\bar{B}^{0} \to \bar{K}^{0} \chi _{c1} $ \textit{etc.} have been measured (for all available values see Table \ref{t0} \cite{29}. The branching ratios of the measured charmed \textit{B} decays range from $\mathcal{O} (10^{-3})\sim \mathcal{ O} (10^{-5} ).$ Therefore, we put our focus mainly to study \textit{B } $\rightarrow$  \textit{PA }decay modes involving charmed axial-vector meson in the final state.

In this work, we investigate the axial-vector emitting decays of bottom (\textit{B}) mesons involving charmed meson states in Cabibbo-Kobayashi-Maskawa (CKM) allowed and suppressed modes. We employ improved Isgur-Scora-Grinstein-Wise (ISGW II) \cite{6} quark model , to evaluate the transition $B\to A/A'$ form factors. It is worth remarking that ISGW II incorporates heavy quark symmetry constraints and hyperfine distortions of the wave functions. For a long time, besides the recent CLF model, ISGW II model has been the only model to give reliable transition form factors from a ground state \textit{s}-wave meson to a low-lying \textit{p}-wave meson. For $B\to P$ transition form factors, we used the BSW quark model \cite{3, 4}. Using the factorization scheme to obtain the decay amplitudes, we calculate the branching ratios of these decay modes. We found that the calculated branching ratios of cabibbo-favored modes such as $B^{-} \to D^{0} a_{1}^{-} $, $B^{-} \to \pi ^{-} \underline{D}_{1}^{0} $ and $B^{-} \to \pi ^{-} D_{1}^{0} $ are in very good agreement with experimental numbers. Another, aim of the present analysis is to discuss the effects of heavy quark symmetry on axial-vector emitting decays involving $D_1$ and $D_{s1}$ mesons. Therefore, we also calculate the $B\to A/A'$ form factors in ISGW II model in HQS constraints. Consequently, we present the analysis of these decays within the heavy quark symmetry constraints. The comparison with experimental observations reveal that the color-suppressed contributions in these decays could be larger than the theoretical estimates. Also, we find some of these decay channels, especially involving $D_{s1} $ states, have large branching ratios comparable to that of the \textit{s}-wave mesons emitting decay modes and are within the reach of future experiments.

The paper is organized as follows. In Sec. II, the meson spectroscopy is discussed. Methodology for calculating $B\to PA$ decays is presented in Sec. III. In Sec. IV, we discuss the numerical results in non-relativistic quark model and in the light of heavy quark symmetry constraints. The summary and conclusions are given in the last section.

  \section{MESON SPECTROSCOPY}

 Both types of axial-vector mesons, ${}^{3} P_{1} ~(J^{PC} =1^{++} $) and ${}^{1} P_{1}~(J^{PC} =1^{+-})$, behave well with respect to the quark model $q\bar{q}$ assignments. Strange and charmed states are most likely a mixture of $^{3} P_{1} $ and $^{1} P_{1} $ states, since there is no quantum number forbidding such mixing. In contrast, diagonal $^{3} P_{1} $ and $^{1} P_{1} $ systems have opposite \textit{C}-parity and cannot mix. Experimentally \cite{29}, the following non-strange and uncharmed mesons have been observed:

  \begin{enumerate}
\item[i.] for ${}^{3} P_{1} $ multiplet, isovector $a_{1} (1.230)$ and two isoscalars $f_{1} (1.285)  $ and $f'_{1} (1.512)$;

  \item[ii.] for ${}^{1} P_{1} $ multiplet, isovector $b_{1} (1.229)$ and two isoscalars $  h_{1} {1.170}$ and $  h'_{1} (1.380)$. \textit{C}-parity of $  h^{'}_{1} (1.380)$ and spin and parity of the $h_{c1} (3.526)$remains to be confirmed. 
\end{enumerate}

Numerical values given in the brackets indicate mass (in GeV) of the respective mesons. In the present analysis, mixing of the isoscalar states of ($1^{++}$) mesons is defined as

\begin{equation} \label{1} 
\begin{array}{l} {f_{1} (1.285)  =\frac{1}{\sqrt{2} } (u\overline{u}+d\overline{d})\cos \phi {}_{A} +(s\overline{s})\sin \phi {}_{A}   ,} \\ {f'_{1} (1.512)  =\frac{1}{\sqrt{2} } (u\overline{u}+d\overline{d})\sin \phi {}_{A} -(s\overline{s})\cos \phi {}_{A} .} \end{array} 
\end{equation} 
\[\chi _{c1} (3.511)=(c\bar{c}),\] 
where
\[\phi _{A} =  \theta (ideal)-\theta _{A} (physical).\] 

Similarly, mixing of two isoscalar mesons $ h_{1} (1.170) $ and $h{'}_{1} (1.380)$ is defined as:

\begin{equation} \label{2} 
\begin{array}{l} {  h_{1} (1.170)=\frac{1}{\sqrt{2} } (u\overline{u}+d\overline{d})\cos \phi {}_{A'} +(s\overline{s})\sin \phi {}_{A'}   ,} \\ {  h'_{1} (1.380)=\frac{1}{\sqrt{2} } (u\overline{u}+d\overline{d})\sin \phi {}_{A'} -(s\overline{s})\cos \phi {}_{A'}   .} \end{array} 
\end{equation} 
\[h_{c1} (3.526)=(c\bar{c}).\] 
Proximity of $a_{1} (1.230)$ and $f_{1} (1.285)  $ and to lesser extent that of $b_{1} (1.229)$ and $  h_{1} (1.170)$ indicates the ideal mixing for both $1^{++} $ and $1^{+-} $ nonets i.e., 
\begin{equation} \label{3} 
\phi _{A}     =    \phi _{A'}     =    0^{\circ } .       
\end{equation} 

States involving a strange quark of $A  (J^{PC} =1^{++} )$ and $A'(J^{PC} =1^{+-} )$mesons mix to generate the physical states in the following manner:
\begin{equation} \label{4} 
\begin{array}{l} {K_{1} (1.270)  =  K_{1A}   \sin \theta _{1}     +  K_{1A'}   \cos \theta _{1} ,} \\ {\underline{K}_{1} (1.400)  =  K_{1A}   \cos \theta _{1}     -  K_{1A'}   \sin \theta _{1} .} \end{array} 
\end{equation} 
where $K_{1A}$ and $K_{1A^{'}} $ denote the strange partners of $a_{1}(1.230)$ and $b_{1}(1.229)$ respectively. Particle Data Group \cite{29} assumes that the mixing is maximal, i.e., $\theta _{1} =45^{\circ } $, whereas $\tau \to K_{1}(1.270)/K_{1} (1.400)+\nu _{\tau } $ data yields $\theta _{1} =\pm   37^{\circ } $ and $\theta _{1} =\pm   58^{\circ } $ \cite{9, 18,19}. Furthermore, the study of $D\to K_{1} (1.270)\pi , ~ K_{1} (1.400)\pi $ decays rules out positive mixing-angle solutions and  $\theta _{1} =-58^{\circ } $ \cite{9} is experimentally favored. However, in a recent phenomenological analysis \cite{30}, it has been shown that the choice of angles for $f - f^{'}$ and  $h - h^{'}$ mixing schemes (that favors ideal mixing) are closely related to the choice of the mixing angle $\theta_1$, therefore,  a mixing angle $\sim 35^{\circ} $ is preferred over $\sim 55^{\circ}$. We use $\theta_1=-37^{\circ} $ in our numerical calculations. 
The mixing of charmed and strange charmed states mesons is similarly given by
\begin{equation} \label{5} 
\begin{array}{l} {D_{1} (2.427)  =  D_{1A}   \sin \theta _{D_{1} }     +  D_{1A'}   \cos \theta _{D_{1} } ,} \\ {\underline{D}_{1} (2.422)  =  D_{1A}   \cos \theta _{D_{1} }     -  D_{1A'}   \sin \theta _{D_{1} } ,} \end{array} 
\end{equation} 
and
\begin{equation} \label{6} 
\begin{array}{l} {D_{s1} (2.460)  =  D_{s1A}   \sin \theta _{D_{s1} }     +  D_{s1A'}   \cos \theta _{D_{s1} } ,} \\ {\underline{D}_{s1} (2.535)  =  D_{s1A}   \cos \theta _{D_{s1} }     -  D_{s1A'}   \sin \theta _{D_{s1} } ,} \end{array} 
\end{equation} 
However, in the heavy quark limit, the physical mass eigenstates with $J^{P} =1^{+} $ are $P_{1}^{{3\mathord{\left/ {\vphantom {3 2}} \right. \kern-\nulldelimiterspace} 2} } $ and $P_{1}^{{1\mathord{\left/ {\vphantom {1 2}} \right. \kern-\nulldelimiterspace} 2} } $ rather than ${}^{3} P_{1} $ and ${}^{1} P_{1} $ states as the heavy quark spin $S_{Q} $ decouples from the other degrees of freedom, so that $S_{Q} $ and the total angular momentum of the light antiquark are separately good quantum numbers \cite{31}. Therefore, we can write
\begin{equation} \label{7} 
\begin{array}{l} {|P_{1}^{{1\mathord{\left/ {\vphantom {1 2}} \right. \kern-\nulldelimiterspace} 2} } >  =  \sqrt{\frac{1}{3} } |{}^{1} P_{1} >-\sqrt{\frac{2}{3} } |{}^{3} P_{1} >,} \\ {} \\ {|P_{1}^{{3\mathord{\left/ {\vphantom {3 2}} \right. \kern-\nulldelimiterspace} 2} } >  =  \sqrt{\frac{2}{3} } |{}^{1} P_{1} >+\sqrt{\frac{1}{3} } |{}^{3} P_{1} >.} \\ {.} \end{array} 
\end{equation} 
Hence, the states $D_{1} (2.427)$ and $\underline{D}_{1} (2.422)$ can be identified with $P_{1}^{{1\mathord{\left/ {\vphantom {1 2}} \right. \kern-\nulldelimiterspace} 2} } $ and $P_{1}^{{3\mathord{\left/ {\vphantom {3 2}} \right. \kern-\nulldelimiterspace} 2} } $, respectively. However, beyond the heavy quark limit, there is a mixing between $P_{1}^{{1\mathord{\left/ {\vphantom {1 2}} \right. \kern-\nulldelimiterspace} 2} } $ and $P_{1}^{{3\mathord{\left/ {\vphantom {3 2}} \right. \kern-\nulldelimiterspace} 2} } $ denoted by 
\begin{equation} \label{8} 
\begin{array}{l} {D_{1} (2.427)  =  D_{1}^{{1\mathord{\left/ {\vphantom {1 2}} \right. \kern-\nulldelimiterspace} 2} } \cos \theta _{2}     +  D_{1}^{{3\mathord{\left/ {\vphantom {3 2}} \right. \kern-\nulldelimiterspace} 2} }   \sin \theta _{2} ,} \\ {\underline{D}_{1} (2.422)  =  -D_{1}^{{1\mathord{\left/ {\vphantom {1 2}} \right. \kern-\nulldelimiterspace} 2} }   \sin \theta _{2}     +D_{1}^{{3\mathord{\left/ {\vphantom {3 2}} \right. \kern-\nulldelimiterspace} 2} } \cos \theta _{2} .} \end{array} 
\end{equation} 
 The mixing angle $\theta _{2} =(-5.7\pm 2.4)^{\circ } $ is obtained by Belle through a detailed   analysis \cite{32}. However, we use a positive mixing angle $\theta _{D_{1} } =17^{\circ } $ based on the study of $D_1(2427)\pi $ production in \textit{B} decays \cite{18}. Likewise for strange axial-vector charmed mesons,

\begin{equation} \label{9} 
\begin{array}{l} {D_{s1} (2.460)  =  D_{s1}^{{1\mathord{\left/ {\vphantom {1 2}} \right. \kern-\nulldelimiterspace} 2} } \cos \theta _{3}     +  D_{s1}^{{3\mathord{\left/ {\vphantom {3 2}} \right. \kern-\nulldelimiterspace} 2} }   \sin \theta _{3} ,} \\ {\underline{D}_{s1} (2.535)  =  -D_{s1}^{{1\mathord{\left/ {\vphantom {1 2}} \right. \kern-\nulldelimiterspace} 2} }   \sin \theta _{3}     +D_{s1}^{{3\mathord{\left/ {\vphantom {3 2}} \right. \kern-\nulldelimiterspace} 2} } \cos \theta _{3} .} \end{array} 
\end{equation} 
$\theta _{3} \approx 7^{\circ } $ is determined from the quark potential model \cite{18, 19, 33}.
For $\eta   $ and $\eta '$ pseudoscalar states, we use 
\begin{equation} \label{10} 
\begin{array}{l} {\eta   (0.547)  =    \frac{1}{\sqrt{2} }   (u\overline{u}+d\overline{d})  \sin \phi _{P} -  (s\overline{s})  \cos \phi _{P} ,} \\ {\eta '  (0.958)  =    \frac{1}{\sqrt{2} }   (u\overline{u}+d\overline{d})  \cos \phi _{P} +  (s\overline{s})\sin \phi _{P} ,} \end{array} 
\end{equation} 
where $\phi _{P} = \theta _{ideal} -\theta _{physical} $, $\theta_{physical}  = -15.4^{\circ}$. $\eta _{c} $ is taken as $\eta _{c} (2.979)=(c\bar{c})$.    

\section{METHODOLOGY}

\subsection{ Weak Hamiltonian}
For bottom changing $\Delta b=1$ decays, the weak Hamiltonian involves the bottom changing current,  
\begin{equation} \label{11} 
J_{\mu } =(\bar{c}b)V_{cb} +(\bar{u}b)V_{ub} ,          
\end{equation} 
where $(\bar{q}_{i} q_{j} )\equiv \bar{q}_{i} \gamma _{\mu } (1-\gamma _{5} )q_{j} $ denotes the weak \textit{V-A }current.  QCD modified weak Hamiltonian is then given below:
\begin{enumerate}
\item[a.] for decays involving $b\to c$ transition,
\end{enumerate}
\ben \label{11a} 
H_{W}^{} =\frac{G_{F} }{\sqrt{2} } \big\{ V_{cb} V_{ud}^{*} [a{}_{1} (\overline{c}b)(\overline{d}u)+a_{2} (\overline{d}b)(\overline{c}u)]+V_{cb} V_{cs}^{*} [a{}_{1} (\overline{c}b)(\overline{s}c)+a_{2} (\overline{s}b)(\overline{c}c)] + \non \\ 
V_{cb} V_{us}^{*} [a{}_{1} (\overline{c}b)(\overline{s}u)+a_{2} (\overline{s}b)(\overline{c}u)]+V_{cb} V_{cd}^{*} [a{}_{1} (\overline{c}b)(\overline{d}c)+a_{2} (\overline{d}b)(\overline{c}c)]\big\}, \en      
\begin{enumerate}
\item[b.]  for decays involving $b\to u$ transition,
\end{enumerate}
\ben \label{11b}{H_{W}^{} =\frac{G_{F} }{\sqrt{2} } \big\{ V_{ub} V_{cs}^{*} [a{}_{1} (\overline{u}b)(\overline{s}c)+a_{2} (\overline{s}b)(\overline{u}c)]+V_{ub} V_{ud}^{*} [a{}_{1} (\overline{u}b)(\overline{d}u)+a_{2} (\overline{d}b)(\overline{u}u)]+} \non\\                                           V_{ub} V_{us}^{*} [a{}_{1} (\overline{u}b)(\overline{s}u)+a_{2} (\overline{s}b)(\overline{u}u)]+                                        V_{ub} V_{cd}^{*} [a{}_{1} (\overline{u}b)(\overline{d}c)+a_{2} (\overline{d}b)(\overline{u}c)]\big\}  \en   

By factorizing matrix elements of the four-quark operator contained in the effective Hamiltonian \eqref{11a} and \eqref{11b}, one can distinguish three classes of decays \cite{31}: 
\begin{enumerate}
\item  The first class contains those decays which can be generated from color singlet current and the decay amplitudes are proportional to $a_{1} $, where $a_{1} (\mu )=c_{1} (\mu )+\frac{1}{N_{c} }   c_{2} (\mu )$, and $N_{c} $ is the number of colors.

\item  A second class of transitions consist of those decays which can be generated from neutral current. The decay amplitude in this class is proportional to $a_{2} $ i.e. for the color suppressed modes  $a_{2} (\mu )=c_{2} (\mu )+\frac{1}{N_{c} }   c_{1} (\mu ).$

\item  The third class of decay modes can be generated from the interference of color singlet and color neutral currents i.e. the $a_{1} $ and $a_{2} $ amplitudes interfere.
\end{enumerate}

However, we follow the convention of large $N_{c} $ limit to fix QCD coefficients $a_{1} \approx c_{1  } $ and $a_{2} \approx c_{2  } $, where \cite{7, 31}:
\begin{center}
$c_{1} (\mu )=1.12$ , $c_{2} (\mu )=-0.26$ at $\mu \approx m_{b}^{2} $.                                     
\end{center}
It may be noted that the decay amplitudes can be expressed as factorizable contributions multiplied by corresponding $a_i$'s that are (renormalization) scale and process independent. As we have mentioned earlier, \textit{B} decays either proceed only via tree diagrams or are tree dominated, therefore we neglect the expected small penguin contributions in our formalism.

\subsection{Decay amplitudes and rates}
The decay rate formula for $B\to PA$ decays is given by
\ben \label{14} 
\Gamma   (B  \to   P  A)    =    \frac{p_{c}^{3} }{8  \pi   m_{A}^{2} }   \left|A(B  \to   P  A)\right|^{2} ,                    
\en 
  where $p_{c} $ is the magnitude of the three-momentum of a final-state particle in the rest frame of $B$ meson and $m_{A} $ denotes the mass of the axial-vector meson.

The factorization scheme expresses the decay amplitudes as a product of the matrix elements of weak currents (up to the weak scale factor of $ \frac{G_{F} }{\sqrt{2} } \times$  CKM elements  $\times$ QCD factor) as
\ben \label{15} 
\left\langle PA \right| H_{w}   \left| B \right\rangle &\sim& \left\langle P \right| J^{\mu } \left| 0 \right\rangle   \left\langle A \right| J_{\mu }   \left| B \right\rangle     +  \left\langle A \right| J^{\mu } \left| 0 \right\rangle \left\langle P \right| J_{\mu } \left| B \right\rangle , \\ \left\langle PA^{'} \right| H_{w}   \left| B \right\rangle   &\sim& \left\langle P \right| J^{\mu } \left| 0 \right\rangle   \left\langle A^{'} \right| J_{\mu} \left| B \right\rangle     +  \left\langle A^{'} \right| J^{\mu } \left| 0 \right\rangle \left\langle P \right| J_{\mu } \left| B \right\rangle . 
\en
Using Lorentz invariance, matrix elements of the current between meson states can be expressed \cite{6} as
\ben \label{16} 
\left\langle P \right| J_{\mu } \left| 0 \right\rangle   &=&  -if_{P} k_{\mu } ,\\ 
\left\langle A \right| J_{\mu } \left| 0 \right\rangle   &=&  \in _{\mu }^{*} m_{A} f_{A} ,\\ 
\left\langle A' \right| J_{\mu } \left| 0 \right\rangle  &=&  \in _{\mu }^{*} m_{A'} f_{A'} ,                    
\en
\ben {\left\langle A(P_{A} ) \right|} J_{\mu } {\left| B(P_{B} ) \right\rangle}   &=&  l\in _{\mu }^{*} +c_{+} (\in ^{*} \cdot P_{B} )(P_{B} +P_{A} )_{\mu } +c_{-} (\in ^{*} \cdot P_{B} )(P_{B} -P_{A} )_{\mu } , \\  {\left\langle A'(P_{A'} ) \right|} J_{\mu } {\left| B(P_{B} ) \right\rangle}  &=&  r\in _{\mu }^{*} +s_{+} (\in ^{*} \cdot P_{B} )(P_{B} +P_{A'} )_{\mu } +s_{-} (\in ^{*} \cdot P_{B} )(P_{B} -P_{A'} )_{\mu } , \en
and
\ben {\left\langle P(P_{P} ) \right|} J_{\mu } {\left| B(P_{B} ) \right\rangle}   &=&  (P_{B \mu } +P_{P \mu } -\frac{m_{B}^{2} -m_{P}^{2} }{q^{2} } q_{\mu } )  F_{1}^{BP} (q^{2} )+\frac{m_{B}^{2} -m_{P}^{2} }{q^{2} } q_{\mu }   F_{0}^{BP} (q^{2} ).\en 
Which yield 
\ben \label{17} 
A(B  \to   PA  )   &=&    (    2    m_{A}     f_{A}   F_{1}^{B\to P} (m_{A}^{2} )      +      f_{P}     F^{B\to A} (m_{P}^{2} )    )  , \\ A(B  \to   PA'  )  &=&    (    2    m_{A'}     f_{A'}   F_{1}^{B\to P} (m_{A'}^{2} )      +      f_{P}     F^{B\to A'} (m_{P}^{2} ) ), \en
where 
\ben \label{18} 
F  ^{B  \to   A}   (m_{P}^{2} )  =  l    +    (m_{B}^{2}     -    m_{A}^{2} )    c_{+}     +    m_{P}^{2}     c_{-}   ,\\                                               F  ^{B\to   A'}   (m_{P}^{2} )  =  r    +    (m_{B}^{2}     -    m_{A'}^{2} )    s_{+}     +    m_{P}^{2}     s_{-}   . 
\en 
\section{DECAY CONSTANT AND FORM FACTORS}
\subsection{Decay constants}
Decay constant of pseudoscalar mesons are well known. In this work, we use the following values of decay constants \cite{9, 34,35} of the pseudoscalar mesons$(0^{-} )$:
\begin{center}
$f_{\pi } $= 0.131 GeV, $f_{K} $= 0.160 GeV,\\ $f_{D} $= 0.208 GeV, $f_{D_{s} } $= 0.273 GeV,\\   $f_{\eta } =0.133$ GeV,             $  f_{\eta '} =  0.126$ GeV and $f_{\eta _{c} } =0.400$ GeV.
\end{center} 
However, for axial-vector meson$(1^{+} )$, decay constants for $J^{PC} =1^{+-} $ mesons may vanish due to the C-parity behavior. Under charge conjunction, the two types of axial-vector mesons transform as
\ben {M_{b}^{a}     (1^{++} )    \to       +M_{a}^{b}   (1^{++} )} \non \\  
 {M_{b}^{a} {\kern 1pt}     (1^{+-} )    \to       -M_{a}^{b} (1^{+-} )}  \en 
where $(a,    b    =  1,    2,    3)$ and $M_{b}^{a} $ denotes meson $3 \times 3$ matrix elements in SU(3) flavor symmetry. Since the weak axial-vector current transforms as $(A_{\mu } )_{b}^{a} \to +(A_{\mu } )_{a}^{b} $ under charge conjunction, only the ($1^{++} $) state can be produced through the axial-vector current in the SU(3) symmetry limit \cite{36}. Particle Data Group \cite{29} assumes that the mixing is maximal, i.e., $\theta =45^{0} ,$ whereas $\tau \to K_{1} (1.270)/K_{1} (1.400)+\nu _{\tau } $ data yields $\theta =\pm   37^{0} $ and $\theta =\pm   58^{0} $. To determine the decay constant of $K_{1} (1.270)$, we use the following formula:
\[\Gamma (\tau \to K_{1} \nu _{\tau } )=\frac{G_{F}^{2} }{16\pi } |V_{us} |^{2} f_{K_{1} }^{2} \frac{(m_{\tau }^{2} +2m_{K_{1} }^{2} )(m_{\tau }^{2} -m_{K_{1} }^{2} )^{2} }{m_{\tau }^{3} } ,               \] 
which gives $f_{K_{1} (1270)} =0.175\pm 0.019$ GeV. The decay constant of $K_{1} (1.400)$ can be obtained from ${f_{K_{1} (1.400)} \mathord{\left/ {\vphantom {f_{K_{1} (1.400)}  f_{K_{1} (1.270)} }} \right. \kern-\nulldelimiterspace} f_{K_{1} (1.270)} } =\cot \theta $. A small value around 0.011 GeV for the decay constant of $K_{1B} $ may arise through SU(3) breaking, which yields $f_{\underline{K}_{1} (1.400)} =f_{K_{1A}   } \cos \theta _{1} -f_{K_{1B}   } \sin \theta _{1}  = -0.232$ GeV for  $\theta_1= -37^{\circ }$ ($-0.087  $ GeV for $\theta _{1} =-58^{\circ } $) \cite{9}. Similarly, decay constant of $a_{1} (1.260) $ can be obtained from $B(\tau \to a_{1} \nu _{\tau } )$. However, this branching ratio is not given in Particle Data Group \cite{29}, although the data on $\tau \to a_{1} \nu _{\tau } \to \rho \pi \nu _{\tau } $ have been reported by various experiments.  We take $f_{a_{1} } =0.203\pm 0.018$ GeV from the analysis given by J.C.R. Bloch \textit{et al.}\cite{37}. For the decay constant of $f_{1}(1.285) $, we assume $f_{f_{1} } \approx f_{a_{1} } $. The decay constants 
\begin{center}
$f_{D_{1A}^{} } =-0.127$ GeV, $f_{D_{1B}^{} } =0.045$ GeV,\\
$f_{D_{s1A}^{} } =-0.121$ GeV, $f_{D_{s1B}^{} } =0.038$ GeV, \\
$f_{\chi _{c1} } \approx -0.207$ GeV.       
\end{center}                   
have been taken from \cite{9}.

\subsection{ \boldmath{$B \to A/A'$} transition form factors in ISGW II quark model}
  We use the improved ISGW II model which describes a more realistic behavior of the form-factor at large momentum transfer \textit{i.e. }($q_{m}^{2} -q^{2} $). In addition to this, the ISGW II model includes various ingredients, such as the heavy quark symmetry constraints, the heavy quark symmetry breaking color-magnetic interaction, relativistic corrections, etc. The form factors have the following expressions in the ISGW II model \cite{6}.
\ben \label{21} 
l &=&-\tilde{m}_{B} \beta _{B} [\frac{1}{\mu _{-} } +\frac{m_{2} \tilde{m}_{A} (\tilde{\omega }-1)}{\beta _{B}^{2} } (\frac{5+\tilde{\omega }}{6m_{1} } -\frac{m_{2} \beta _{B}^{2} }{2\mu _{-} \beta _{BA}^{2} } )]  F_{5}^{(l)} ,\non \\ \non
 c_{+} +c_{-} &=&-\frac{m_{2} \tilde{m}_{A} }{2m_{1} \tilde{m}_{B} \beta _{B}^{} } \left(1-\frac{m_{1} m_{2} \beta _{B}^{2} }{2\tilde{m}_{A} \mu _{-} \beta _{BA}^{2} } \right)F^{(c_{+} +c_{-} )} , \\ \non c_{+} -c_{-} &=&-\frac{m_{2} \tilde{m}_{A} }{2m_{1} \tilde{m}_{B} \beta _{B}^{} } \left(\frac{\tilde{\omega }+2}{3} -\frac{m_{1} m_{2} \beta _{B}^{2} }{2\tilde{m}_{A} \mu _{-} \beta _{BA}^{2} } \right)F^{(c_{+} -c_{-} )},\non  \\ 
 r&=&\frac{\tilde{m}_{B} \beta _{B} }{\sqrt{2} } [\frac{1}{\mu _{+} } +\frac{m_{2} \tilde{m}_{A} }{3m_{1} \beta _{B}^{2} } (\tilde{\omega }-1)^{2} ]  F_{5}^{(r)} ,  \\ \non
s_{+} +s_{-} &=&-\frac{m_{2} }{2\tilde{m}_{B} \beta _{B}^{} } \left(1-\frac{m_{2} }{m_{1} } +\frac{m_{2} \beta _{B}^{2} }{2\mu _{+} \beta _{BA}^{2} } \right)F^{(s_{+} +s_{-} )} , \\ \non s_{+} -s_{-} &=&-\frac{m_{2} }{2m_{1} \beta _{B}^{} } \left(\frac{4-\tilde{\omega }}{3} -\frac{m_{1} m_{2} \beta _{B}^{2} }{2\tilde{m}_{A} \mu _{+} \beta _{BA}^{2} } \right)F^{(s_{+} -s_{-} )} , \en 
where
\ben \label{22} 
F_{5}^{(l)} &=& F_{5}^{(r)} =F_{5} (\frac{\bar{m}_{B} }{\tilde{m}_{B} } )^{1/2} (\frac{\bar{m}_{A} }{\tilde{m}_{A} } )^{1/2} , \non \\  F_{5}^{(c_{+} +c_{-} )} &=& F_{5}^{(s_{+} +s_{-} )} =F_{5} (\frac{\bar{m}_{B} }{\tilde{m}_{B} } )^{-3/2} (\frac{\bar{m}_{A} }{\tilde{m}_{A} } )^{1/2} , \\  F_{5}^{(c_{+} -c_{-} )} &=& F_{5}^{(s_{+} -s_{-} )} = F_{5} (\frac{\bar{m}_{B} }{\tilde{m}_{B} } )^{-1/2} (\frac{\bar{m}_{A} }{\tilde{m}_{A} } )^{-1/2}. 
\non \en 
The $t(\equiv q^{2} )$dependence is given by
\begin{equation} \label{24} 
\tilde{\omega }-1=\frac{t_{m} -t}{2\bar{m}_{B} \bar{m}_{A} } ,     
\end{equation} 
and 
\begin{equation} \label{25} 
F_{5} =\left(\frac{\tilde{m}_{A} }{\tilde{m}_{B} } \right)^{{\raise0.7ex\hbox{$ 1 $}\!\mathord{\left/ {\vphantom {1 2}} \right. \kern-\nulldelimiterspace}\!\lower0.7ex\hbox{$ 2 $}} } \left(\frac{\beta _{B} \beta _{A} }{B_{BA} } \right)^{{\raise0.7ex\hbox{$ 5 $}\!\mathord{\left/ {\vphantom {5 2}} \right. \kern-\nulldelimiterspace}\!\lower0.7ex\hbox{$ 2 $}} } \left[1+\frac{1}{18} h^{2} (t_{m} -t)\right]^{-3} ,                
\end{equation} 
where
\[h^{2}   =  \frac{3}{4m_{c} m_{q} } +\frac{3m_{d}^{2} }{2\bar{m}_{B} \bar{m}_{A} \beta _{BA}^{2} } +\frac{1}{\bar{m}_{B} \bar{m}_{A} } (\frac{16}{33-2n_{f} } )\ln [\frac{\alpha _{S} (\mu _{QM} )}{\alpha _{S} (m_{q} )} ],            \] 
with 
\begin{equation} \label{26} 
\beta _{BA}^{2} =\frac{1}{2}     \left(\beta _{B}^{2} +\beta _{A}^{2} \right),                       
\end{equation} 
and 
\[\mu _{\pm } =\left(\frac{1}{m_{q} } \pm \frac{1}{m_{b} } \right)^{-1} .\] 
$\tilde{m}$ is the sum of the mesons constituent quarks masses, $\bar{m}$ is the hyperfine averaged physical masses, $n_f$ is the number of active flavors, which is taken to be five in the present case, $t_{m} =(m_{B} -m_{A} )^{2} $ is the maximum momentum transfer and $\mu _{QM} $ is the quark model scale. The subscript in the \textit{q} depends upon the quark currents $\bar{q}\gamma _{\mu } b$ and $\bar{q}\gamma _{\mu } \gamma _{5} b$ appearing in different transitions. The values of parameter $\beta $ for different \textit{s}-wave and \textit{p}-wave mesons are given in the Table \ref{t1}.  We use the following constituent quark masses 

\begin{center}
$m_{u} =m_{d} =0.31,~ m_{s} =0.49, ~m_{c} =1.7$, and $    m_{b} =5.0$, 
\end{center} 
  to calculate the form factors for $B\to A$and $B\to A'$ transitions. The obtained form factors are given in Tables \ref{t2} and \ref{t4}.
For $B\to P$ transition, we use the well-established BSW \cite{3, 4} quark model. Using constituent quark masses same as for ISGW model and the average transverse quark momentum inside a meson\textit{ $\omega$} = 0.5, we obtain the $B\to P$ transition form factors as shown in column 2 of Table \ref{t4}. 
                                                                               
\section{NUMERICAL RESULTS AND DISCUSSIONS }
Sandwiching the weak Hamiltonian \eqref{11a} \& \eqref{11b} between the initial and final states, we obtain the decay amplitudes of $B^{-} $, $\bar{B}^{0} $ mesons for the various decay modes as given in the Tables \ref{t5}, \ref{t6}, \ref{t7}, \ref{t8}, \ref{t9} and \ref{t10}.  Finally, the calculated branching ratios are given (which are expected to be tree dominated) in CKM-favored and CKM-suppressed modes involving $b\to c$ and $b\to u$ transitions. 

The results are given in Tables \ref{t11}, \ref{t12}, \ref{t13}, \ref{t14}, \ref{t15} and \ref{t16} for various possible modes. We also calculated the form factors and branching ratios for decays involving charm meson in the light of heavy quark symmetry constraints given in Tables \ref{t17} and \ref{t18}. The following are our results,

\subsection{\boldmath{$B\to PA$ decays involving $b\to c$ transition}}

\subsubsection{${\rm \Delta }b=1,  {\rm \Delta }C=1,{\rm \Delta }S=0$ mode: }
$\bar{B}^{0} \to D^{+} a_{1}^{-} $, $B^{-} \to D^{0} a_{1}^{-} $ and $B^{-} \to \pi ^{-} \underline{D}_{1}^{0} $ are the dominant decays with branching ratios of the order of $\mathcal{O}(10^{-2}) \sim \mathcal{O}(10^{-3})$. The highest branching ratio is for $\bar{B}^{0} \to D^{+} a_{1}^{-} $ decay. Four of the decay channels are experimentally measured which are discussed as follows: 

\begin{enumerate}
\item [i.] For $\bar{B}\to \pi D_{1} $ decay mode, 
 \begin{flushleft}
$B(B^{-} \to \pi ^{-} \underline{D}_{1}^{0} )= 1.4\times10^{-3} ~~~~~~~~~~~~~~~~~~(1.5\pm 0.6)\times 10^{-3} $ (\textit{Exp});
$B(B^{-} \to \pi ^{-} D_{1}^{0} ) = 8.3\times10^{-4} ~~~~~~~~~~~~~~~~~~(7.5\pm 1.7)\times 10^{-4} $\cite{32} (\textit{Exp}).     
\end{flushleft}

  The calculated branching ratios are in good agreement with the available experimental results. The difference of roughly a factor of 2 between branching ratios of $B^{-} \to \pi ^{-} \underline{D}_{1}^{0} $ and $B^{-} \to \pi ^{-} D_{1}^{0} $ can be attributed to the constructive and destructive interference between color-favored and color-suppressed transitions. The $B^{-} \to \pi ^{-} D_{1}^{0} $ amplitude receives contribution from destructive interference between color-allowed and color-suppressed currents and hence, has smaller branching ratio as compared to as the $B^{-} \to \pi ^{-} \underline{D}_{1}^{0} $ decay. On the other hand, branching ratios of the color-allowed decays resulting from internal W-emission tree processes are $B(\bar{B}^{0} \to \pi ^{-} \underline{D}_{1}^{+} )$\textbf{=} $2.2\times10^{-3}$ and $B(\bar{B}^{0} \to \pi ^{-} D_{1}^{+} )=8.3\times10^{-4}$.

\item[ii.]  For $\bar{B}\to Da$ decay mode the calculated branching ratios are
\begin{flushleft}
 $B(\bar{B}^{0} \to D^{+} a_{1}^{-} )=1.1 \times 10^{-2}~~~~~~~~~~~~~~~~~~   (0.60\pm 0.33)\times 10^{-2} $ (\textit{Exp}) $B(B^{-} \to D^{0} a_{1}^{-} )=5.5\times10^{-3}~~~~~~~~~~~~~~~~~~                 (4\pm 4)\times 10^{-3} $ (\textit{Exp}).
\end{flushleft}
Both of the calculated decay modes are consistent with experimental results within the error. The $\bar{B}^{0} \to D^{+} a_{1}^{-} $ decay receive contribution from color-favored transition only, however, the $B^{-} \to D^{0} a_{1}^{-} $decay get contributions through the destructive interference between color-favored and color-suppressed transitions resulting in smaller branching ratio. The next order color-suppressed decays:  $\bar{B}^{0} \to D^{0} a_{1}^{0} /D^{0} f_{1} /D^{0} b_{1}^{0} /D^{0} h_{1} $ have branching ratios of the $\mathcal{O}(10^{-4})$.

\item[iii.]  Decays $\bar{B}^{0} \to D_{s}^{+} K_{1}^{-} /D_{s}^{+} \underline{K}_{1}^{-} /K^{-} D_{s1}^{+} /K^{-} \underline{D}_{s1}^{+} $  are forbidden in the spectator model. These decays may be generated through quark annihilation diagrams.  However, these annihilation contributions involve creation of $(s\bar{s})$ pair which is relatively suppressed. $\bar{B}^{0} \to D^{0} f'_{1} /D^{0} h'_{1} $ are forbidden in the limit of ideal mixing for $f_{1} -f'_{1} $ and $h_{1} -h'_{1} $ mesons. Any deviation from the ideal mixing may generate these decays. It may be noted that no penguin or single quark transition contribute to this decay mode. However, $\bar{B}^{0} $ meson decays of this mode may have contribution from annihilation diagrams. 
\end{enumerate}

\subsubsection{${\rm \Delta }b=1,  {\rm \Delta }C=0,{\rm \Delta }S=-1$ mode: }

\begin{enumerate}
\item [i.] We obtain $B(B^{-} \to D^{0} D_{s1}^{-} )= 2.0\times10^{-3},~B(\bar{B}^{0} \to D_{s}^{-} D_{1}^{+} ) = 2.0 \times 10^{-3},~B(B^{-} \to D_{s}^{-} D_{1}^{0} ) = 2.1 \times 10^{-3}, ~ B(\bar{B}^{0} \to D_{s}^{-} \underline{D}_{1}^{+} ) = 3.7 \times 10^{-3},~ B(B^{-} \to D_{s}^{-} \underline{D}_{1}^{0} )  =  3.9 \times 10^{-3}$, and $B(\bar{B}^{0} \to D^{+} D_{s1}^{-} )= 1.9 \times 10^{-3}$. In spite of the kinematic suppression, these modes acquire large branching ratios as these involve color-favored quark diagram and large value of decay constants of the charmed mesons.

\item [ii.]  For color-suppressed $K\chi _{c1} $ mode, we obtain $B(B^{-} \to K^{-} \chi _{c1} )=1.3\times {\rm 1}0^{-{\rm 4}} $, $B(\bar{B}^{0} \to \bar{K}^{0} \chi _{c1} )=1.2\times {\rm 1}0^{-{\rm 4}} $, which are smaller than the measured experimental branching ratios, \textit{i.e.} $(4.79\pm 0.23)\times 10^{-4} $ and $(3.9\pm 0.4)\times 10^{-4} $, respectively, by a factor of $\sim 3.5$. Though, it may be remarked that penguin and annihilation diagrams do not contribute to these decays.

\item[iii.]  Due to the vanishing decay constant $(f_{A'} )$, decays $B^{-} \to K^{-} h_{c1} $ and $\bar{B}^{0} \to \bar{K}^{0} h_{c1} $ are forbidden in the present analysis. Annihilation diagrams do not contribute to this decay mode. However, $\bar{B}^{0} \to DD_{s1}^{} $$/D\underline{D}_{s1}^{} $$/D_{s}^{} D_{1}^{} /D_{s}^{} \underline{D}_{1}^{} $ decay modes may have suppressed contribution from penguin diagrams which include $(c\bar{c})$ pair. 
\end{enumerate}

\subsubsection{${\rm \Delta }b=1,  {\rm \Delta }C=0,{\rm \Delta }S=0$ mode: }
\begin{enumerate}

\item[i.] For dominant decay, we predict branching ratios for $B^{-} \to D^{0} \underline{D}_{1}^{-} /D^{-} \underline{D}_{1}^{0} $ and$\bar{B}^{0} \to D^{-} \underline{D}_{1}^{+} /D^{+} \underline{D}_{1}^{-} $  of the $\mathcal{O}(10^{-4})$. 

\item[ii.]  In the present analysis, we obtain $B(B^{-} \to \pi ^{-} \chi _{c1} )= 0.6 \times 10^{-5}$ and $B(\bar{B}^{0} \to \pi ^{0} \chi _{c1} ) = 0.3 \times 10^{-5}$ which are smaller than the experimental branching ratio $(2.2\pm 0.6)\times 10^{-5} $ and $(1.12\pm 0.28)\times 10^{-5} $, respectively.

\item[iii.]  $\bar{B}^{0} \to \pi ^{0} h_{c1} /\eta 'h_{c1} /D^{0} \bar{D}_{1}^{0} /D^{0} \underline{\bar{D}}_{1}^{0} /D_{s}^{+} D_{s1}^{-} /D_{s}^{+} \underline{D}_{s1}^{-} /\bar{D}^{0} D_{1}^{0}   /\bar{D}^{0} \underline{D}_{1}^{0}   /D_{s}^{-} D_{s1}^{+} /D_{s}^{-} \underline{D}_{s1}^{+} /\eta _{c} f'_{1} /$ $\eta _{c} h'_{1}$ decays are forbidden in the present analysis. Annihilation diagrams, elastic FSI and penguin diagrams may generate these decays to the naked charm mesons. However, decays emitting charmonium $h_{c1} $ remains forbidden in the ideal mixing limit.
\end{enumerate}

\subsubsection{${\rm \Delta }b=1,  {\rm \Delta }C=1,{\rm \Delta }S=-1$ mode: }
Branching ratios of the dominant decays in the present mode are $B(B^{-} \to K^{-} \underline{D}_{1}^{0} ) = 1.1 \times 10^{-4}, ~B(\bar{B}^{0} \to D^{+} \underline{K}_{1}^{-} )  = 1.3 \times 10^{-4}, ~B(B^{-} \to D^{0} K_{1}^{-} ) = 1.6\times10^{-4}, ~B(\bar{B}^{0} \to K^{-} \underline{D}_{1}^{+} )  = 1.6 \times 10^{-4}, ~B(B^{-} \to D^{0} \underline{K}_{1}^{-} ) = 1.7 \times 10^{-4}$, and $B(\bar{B}^{0} \to D^{+} K_{1}^{-} ) = 4.5 \times 10^{-4}$.

\subsection{\boldmath{$B\to PA$ decays involving $b\to u$ transition}}

\subsubsection{$\Delta b=1,  \Delta C=-1,\Delta S=-1$ mode: }
\begin{enumerate}

\item[i.] Dominant decays in the present mode are $B(B^{-} \to D_{s}^{-} f_{1} ) =  1.3 \times 10^{-4}, ~B(B^{-} \to D_{s}^{-} a_{1}^{0} ) =  1.4 \times 10^{-4},~B(\bar{B}^{0} \to D_{s}^{-} b_{1}^{+} ) = 1.4 \times 10^{-4}$, and $ B(\bar{B}^{0} \to D_{s}^{-} a_{1}^{+} )=2.7 \times 10^{-4}$.

\item [ii.] Calculated branching ratios $B(B^{-} \to D_{s}^{-} a_{1}^{0} )=1.4 \times 10^{-4},$ and   $B(\bar{B}^{0} \to D_{s}^{-} a_{1}^{+} )  = 2.7 \times 10^{-4}$ are consistent with the experimental upper limits $<1.8\times 10^{-3} $ and $<2.2\times 10^{-3} $. 

\item[iii.]  Decays $B^{-} \to \bar{K}^{0} D_{1}^{-} /\bar{K}^{0} \underline{D}_{1}^{-} /D^{-} \bar{K}_{1}^{0} /D^{-} \underline{\bar{K}}_{1}^{0} /D_{s}^{-} f'_{1} /D_{s}^{-} h'_{1} $ are forbidden in the present analysis. Annihilation and FSIs may generate these decays.
\item[iv.] Decay channels in $\Delta b=1,  \Delta C=-1,\Delta S=0$ mode are highly suppressed with branching ratios of $ \mathcal{O}(10^{-6})\sim \mathcal{O}(10^{-11})$.
\end{enumerate}

\subsection{ Based on Heavy Quark Symmetry Constraints:}

To compare our results with the exiting theoretical analyses, we calculate the branching ratios of \textit{B} mesons decays involving charm meson in light of heavy quark symmetry constraints. We employ the ISGW II quark model, which follows heavy quark symmetry, to evaluate the form factors involved in the following transitions: 

\[\begin{array}{l} {{\left\langle D^{3/2} (P_{A} ) \right|} J_{\mu } {\left| B(P_{B} ) \right\rangle}   =  l_{3/2} \in _{\mu }^{*} +c_{+}^{3/2} (\in ^{*} \cdot P_{B} )(P_{B} +P_{A} )_{\mu } +c_{-}^{3/2} (\in ^{*} \cdot P_{B} )(P_{B} -P_{A} )_{\mu } ,} \\ {{\left\langle D^{1/2} (P_{A'} ) \right|} J_{\mu } {\left| B(P_{B} ) \right\rangle}   =  l_{1/2} \in _{\mu }^{*} +c_{+}^{1/2} (\in ^{*} \cdot P_{B} )(P_{B} +P_{A} )_{\mu } +c_{-}^{1/2} (\in ^{*} \cdot P_{B} )(P_{B} -P_{A} )_{\mu } .} \end{array}\] 
    We have used HQS constraints (for relations see \cite{4}) to calculate the form factors which are given in rows 7 and 8 of Table II. It may be noted that the signs of the calculated form factors are consistent with heavy quark expectations. Using these form factor and decay constant values, $f_{D_{1}^{1/2} } =0.177$ GeV, and $f_{D_{1}^{3/2} } =-0.045$ GeV \cite{35}, we calculate the branching ratios: $B(B^{-} \to \pi ^{-} \underline{D}_{1}^{0} )  =  1.4 \times 10^{-3}$, and $B(B^{-} \to \pi ^{-} D_{1}^{0} )  = 2.7 \times 10^{-3}$. 

The branching ratio for $B^{-} \to \pi ^{-} \underline{D}_{1}^{0} $ is consistent with the experiment, however, the branching ratio of $B^{-} \to \pi ^{-} D_{1}^{0} $ becomes larger by a factor of $\sim 2$ in the light of HQS constraints. Similar observations could be made for $B(\bar{B}^{0} \to \pi ^{-} D_{1}^{+} ) =  3.2 \times 10^{-3}$ and $B(\bar{B}^{0} \to \pi ^{-} \underline{D}_{1}^{+} )= 5.0 \times 10^{-4}$. We wish to point out that $\bar{B}^{0} \to \pi \underline{D}_{1}$ decays receive contribution from color-favored diagrams only, and result for $B(\bar{B}^{0} \to \pi ^{-} \underline{D}_{1}^{+} )$ decay is consistent with the experimental average \cite{38}.
 
\begin{figure}[h]
\centering
\includegraphics[width=0.6\textwidth]{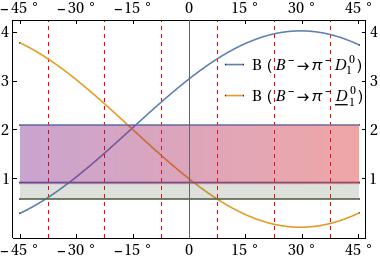}
\caption{Plot of branching ratios vs mixing angle. Experimental branching ratio ranges are shown as shaded regions: brighter for $B(B^{-} \to \pi ^{-} D_{1}^{0} )$\textbf{ }and darker for $B(B^{-} \to \pi ^{-} \underline{D}_{1}^{0} )$}
\label{fig1}
\end{figure}

  We use the experimentally determined $D_{1}^{1/2} -D_{1}^{3/2} $mixing angle $-5.7^{\circ } $\cite{32} to calculate the branching ratios and compare those values with other works. FIG.\ref{fig1}, shows the variation of the branching ratios $B(B^{-} \to \pi ^{-} \underline{D}_{1}^{0} )$ and $B(B^{-} \to \pi ^{-} D_{1}^{0} )$ decays\textbf{ }w.r.t. mixing angle, $\theta _{D} $, that supports our choice of negative mixing angle (to be consistent with experimental values). 

Aforementioned, the decay modes $B^{-} \to \pi ^{-} D_{1}^{0} $ and  $B^{-} \to \pi ^{-} \underline{D}_{1}^{0} $ belong to class III decays, which receive contributions from constructive and destructive interference of color-favored and color-suppressed transitions. The experimental results are pointing to the fact that these contributions cannot be ignored. We wish to remark that, in heavy quark limit \cite{31}, the contributions from color-suppressed amplitudes are further suppressed by a factor of 1/$m_Q$. However, the inconsistency with experiment in case of $B^{-} \to \pi ^{-} D_{1}^{0} $ decay indicates large contribution from color-suppressed amplitude. Also, the implications of HQS framework are such that the theoretical results would become merely independent of color-suppressed contributions. Therefore, yielding the relations: 
 \[B(B^{-} \to \pi ^{-} \underline{D}_{1}^{0} )=B(\bar{B}^{0} \to \pi ^{-} \underline{D}_{1}^{+} );\] 
\[ B(B^{-} \to \pi ^{-} D_{1}^{0} )=B(\bar{B}^{0} \to \pi ^{-} D_{1}^{+} ),\] 
in heavy quark limit. However, these relations may not be satisfied in some cases because, relatively, large contribution from color-suppressed diagrams is expected by experiment. 

 \begin{figure}[h]
\centering
\includegraphics[width=0.6\textwidth]{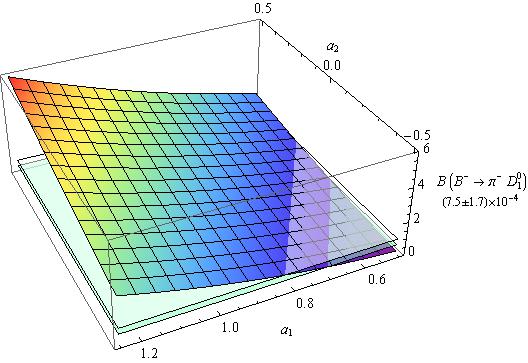}
\caption{Plot of theoretical branching ratio of $B^{-} \to \pi ^{-} D_{1}^{0} $ decay w.r.t. parameters $a_1$ and $a_2$ in heavy quark symmetry constraints. The intersecting parallel planes represent the upper and lower limits of experimental branching ratio.}
\label{fig2}
\end{figure}

\begin{figure}[h]
\centering
\includegraphics[width=0.6\textwidth]{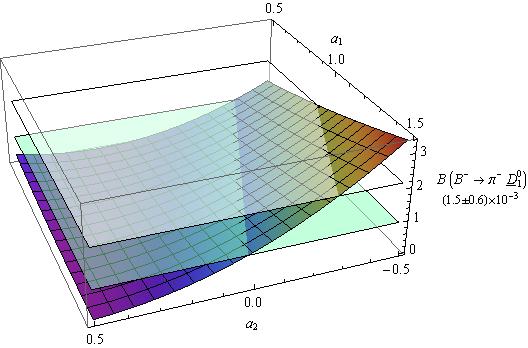}
\caption{Plot of branching ratio $B^{-} \to \pi ^{-} \underline{D}_{1}^{0} $ decay w.r.t. parameters  $a_1$ and  $a_2$ in heavy quark symmetry constraints. The intersecting parallel planes represent the upper and lower limits of experimental branching ratio.}
\label{fig3}
\end{figure}
 In order to get a clearer picture, we plotted the theoretical branching ratios of $B^{-} \to \pi ^{-} D_{1}^{0} $ and $B^{-} \to \pi ^{-} \underline{D}_{1}^{0}$ decays w.r.t. parameters  $a_1$ and  $a_2$ as shown in FIGs. \ref{fig2} and \ref{fig3}. These plots show that the experimental branching ratio for $B^{-} \to \pi ^{-} D_{1}^{0} $ decay support larger magnitude (with negative sign) for color-suppressed amplitude proportional to parameter $a_2$ and relatively, smaller magnitude for color-favored amplitude proportional to $a_1$. It is interesting to note that the same choice of parameters $a_1$ and $a_2$  can also be used in case of $B^{-} \to \pi ^{-} \underline{D}_{1}^{0} $ decay for which experimental branching ratio shows a large overlap region with respect to color-suppressed and color-favored contributions. Therefore, a choice of $\left|{a_{2} \mathord{\left/ {\vphantom {a_{2}  a_{1} }} \right. \kern-\nulldelimiterspace} a_{1} } \right|=0.58$, that fits the experimental observations, clearly indicates the larger contributions from color-suppressed transitions. Once taken in to account, the smaller magnitude of color-favored class I transitions would bring $\bar{B}\to \pi D_{1} $ decays closer to the theoretical expectations \cite{18, 19, 21}. 

In order to compare our results with the available theoretical works \cite{18, 19, 21} in heavy quark limit, we listed their results in column 3 of Table XVII. These analyses are mainly focused on CKM-favored decays \textit{e.g.} $B(B^{-} \to \pi ^{-} \underline{D}_{1}^{0}  ) =1.1 \times 10^{-3}, ~B(\bar{B}^{0} \to \pi ^{-} \underline{D}_{1}^{+} ) =1.1 \times 10^{-3},~ B(\bar{B}^{0} \to \pi ^{-} D_{1}^{+} ) =1.5 \times 10^{-4},$ and $B(B^{-} \to \pi ^{-} D_{1}^{0} ) = 3.7 \times10^{-4}$. The present value for $B^{-} \to \pi ^{-} \underline{D}_{1}^{0} $ decay mode is consistent with their theoretical result. Note that the inconsistencies, in comparison, arise due to the difference of form-factors (owing to the different constituent quark masses), decay constants and their signs; and the choice of sign for the mixing angle \cite{18, 19}. It has been pointed out, in CLF approach, result for $B(B^{-} \to \pi ^{-} D_{1}^{0} )$ could only be explained if a positive sign is taken for $f_{D_{1}^{3/2} } $ decay constant, which is not the case in our results. It is worth mentioning that CLF approach \cite{19} also supports a large contribution from color-suppressed amplitudes. Moreover, Jugeau \textit{et al}. \cite{21} and Cheng \textit{et al}. \cite{19} have used experimental branching for $B^{-} \to \pi ^{-} D_{1}^{0} $ to estimate the decay constants and form factors. It could be seen that heavy quark corrections may result in the large deviations from theoretical expectations in the present scenario. 

  We wish to emphasize that $a_2$ parameter cannot be calculated in the QCD factorization for $\bar{B}\to D^{**} \pi $ type decays because \textit{D} meson being heavy and slow cannot be decoupled from ($B\pi $) system. Thus, soft interaction between ($B\pi $) system and the charm quark of \textit{D} meson will be considerably different from the interaction between the ($B\pi $) system and light spectator quark of \textit{D} meson \cite{39, 40}. This indicates that the nonfactorizable contributions to color-suppressed transitions will be dominated by nonperturbative effects. On the other hand, in soft collinear effective theory (SCET), unlike naive $a_2$ factorization, the type II decays are shown to be factorizable into a pion light-cone wave function and a $\bar{B}{\rm \; }\to {\rm \; }D_{}^{*} $ soft distribution function \cite{17}. Later, they have extended their formalism to color-allowed and color-suppressed $\bar{B}{\rm \; }\to {\rm \; }D_{1} M$  and $\bar{B}{\rm \; }\to {\rm \; }D_{2}^{*} M$ decays in the light of HQS constraints and give the relations for branching fractions in leading order (for equal strong phases in both channels):
\begin{equation*}
\frac{B(\bar{B}\to D_{1} \pi )}{B(\bar{B} \to D_{2}^{*} \pi )} =1; ~~~~~~~(0.54\pm 0.18)(Exp).
\end{equation*}

   Clearly, this equality do not compete well with the existing experimental observation of $0.54\pm 0.18$ by Belle \cite{41}. Thus, the discrepancy between experimental and theoretical expectations in various formalisms is evident. Lastly, we list our results for CKM-favored and CKM-suppressed modes in the Table XVII and Table XVIII, respectively, for comparison with other works.
\section{SUMMARY AND CONCLUSIONS}
  In this paper, we have studied the hadronic weak decays of bottom mesons emitting pseudoscalar and axial-vector mesons. We have employed ISGW II \cite{6} model to determine the $B\to A/A'$ transition form factors in, both, non-relativistic quark model framework and heavy quark symmetry constraints. Consequently, the decay amplitudes and the branching ratios of  $B\to PA$ decays involving $b\to c$ and $b\to u$ transitions in CKM-favored and CKM-suppressed modes are obtained. We draw the following observations:

\begin{enumerate}
\item  Aforementioned, we apply the non-relativistic framework to determine the form factors and the branching ratios for $\bar{B}\to \pi D_{1} /Da_{1} $ decay modes. The branching ratios for these modes are of the order $10^{-2} \sim 10^{-4}$ which are in good agreement with experimental results at $\theta _{D} =17^{\circ } $.  It is interesting to note that the theoretical expectations favor the positive sign of mixing angle when compared with experimental results. 

\item  Though kinematically suppressed, the color-favored $\bar{B}\to DD_{s1}^{} $ and $\bar{B}\to D_{s}^{} D_{1}^{} $ modes have larger branching ratios of the order $10^{-3} \sim 10^{-4}$. The branching ratios for color-suppressed $\bar{B}\to K\chi _{c1} $ decays are of the same order as observed ones but smaller in magnitude.

\item  In CKM-suppressed, ${\rm \Delta }b=1,  {\rm \Delta }C=0,{\rm \Delta }S=0$ and ${\rm \Delta }b=1,  {\rm \Delta }C=1,{\rm \Delta }S=-1$, modes few branching ratios are of the order of $10^{-4}$ which are well within the reach of present experiments.
\end{enumerate}

  Also, we analysed the charm axial-vector meson emitting decays in ISGW II quark model in heavy quark symmetry constraints. We obtained the relevant form factors and branching ratios in CKM-favored mode.

\begin{enumerate}
\item  We calculate the branching ratios for class III type  $\bar{B}\to \pi D_{1} $ decays. The \textit{B}($B^{-} \to \pi ^{-} \underline{D}_{1}^{0} $) decay is consistent with the experimental number, however the \textit{B}($B^{-} \to \pi ^{-} D_{1}^{0} $) decay is larger than experimental expectations. Although, the color-suppressed amplitudes are further supressed by a factor of 1/$m_Q$ in heavy quark limit, the experimental values requires a larger contribution from color-suppressed transitions. Therefore, the branching relations obtained in heavy quark symmetry expectations: $B(B^{-} \to \pi ^{-} \underline{D}_{1}^{0} )=B(\bar{B}^{0} \to \pi ^{-} \underline{D}_{1}^{+} )$ and $B(B^{-} \to \pi ^{-} D_{1}^{0} )=B(\bar{B}^{0} \to \pi ^{-} D_{1}^{+} )$ may not be satisfied.  

\item  Furthermore, the analysis of $\bar{B}\to \pi D_{1} $ decay channels yield the choice of $\left|{a_{2} \mathord{\left/ {\vphantom {a_{2}  a_{1} }} \right. \kern-\nulldelimiterspace} a_{1} } \right|=0.58$, which fits the experimental observation, indicating the need of large color-suppressed amplitude. Thus, the large magnitude for color-suppressed amplitude, $a_2$, and a relatively smaller magnitude for color-favored amplitude, $a_1$, aid the experimental branching results for $B^{-} \to \pi ^{-} D_{1}^{0} $ and $B^{-} \to \pi ^{-} \underline{D}_{1}^{0} $ decay channels. It may also be pointed out that the smaller magnitude of color-favored class I contributions bring our results closer to the other theoretical expectations. 
\end{enumerate}

  Thus, the understanding of class III decays which receive contribution from constructive and destructive interference of color-favored and color-suppressed transitions, is of utmost importance to resolve the puzzle of larger magnitude of $a_2$. More precise experimental information on such decays will help theory to access the nonfactorizable contributions in these processes. 

  ~~
\newpage

\begin{table}
\centering
\caption{Experimentally measured Branching Ratios for $B\to PA$ decays}
\label{t0}

\begin{tabular}{|c|c|} \hline 
\textbf{Mode} & \textbf{Experimental Branchings} \\ \hline 
$B^{-} \to D^{0} a_{1}^{-} $ & $(4\pm 4)\times 10^{-3} $ \\ \hline 
$B^{-} \to \pi ^{-} \underline{D}_{1}^{0} $ & $(1.5\pm 0.6)\times 10^{-3} $ \\ \hline 
$B^{-} \to \pi ^{-} D_{1}^{0} $ & (7.5 $\pm$ 1.7) $\times 10^{-4}$\cite{32} \\ \hline 
$B^{-} \to \pi ^{-} \chi _{c1} $ & $(2.2\pm 0.5)\times 10^{-5} $ \\ \hline 
$B^{-} \to K^{-} \chi _{c1} $ & $(4.79\pm 0.23)\times 10^{-4} $ \\ \hline 
$B^{-} \to \bar{K}^{0} a_{1}^{-} $ & $(3.5\pm 0.7)\times 10^{-5} $ \\ \hline 
$B^{-} \to \pi ^{0} a_{1}^{-} $ & $(2.6\pm 0.7)\times 10^{-5} $ \\ \hline 
$B^{-} \to \pi ^{-} a_{1}^{0} $ & $(2.0\pm 0.6)\times 10^{-5} $ \\ \hline 
$\bar{B}^{0} \to D^{+} a_{1}^{-} $ & $(6.0\pm 3.3)\times 10^{-3} $ \\ \hline 
$\bar{B}^{0} \to \chi _{c1} \pi ^{0} $ & $(1.12\pm 0.28)\times 10^{-5} $ \\ \hline 
$\bar{B}^{0} \to \bar{K}^{0} \chi _{c1} $ & $(3.9\pm 0.4)\times 10^{-4} $ \\ \hline 
$\bar{B}^{0} \to K^{-} a_{1}^{+} $ & $(1.6\pm 0.4)\times 10^{-5} $ \\ \hline 
$\bar{B}^{0} \to \pi ^{\mp } a_{1}^{\pm } $ & $(2.6\pm 0.5)\times 10^{-5} $ \\ \hline 
$B^{-} \to D_{s}^{-} a_{1}^{0} $ & $<1.8\times 10^{-3} $ \\ \hline 
$B^{-} \to \pi ^{-} \bar{K}_{1}^{0} (1270)$ & $<4.0\times 10^{-5} $ \\ \hline 
$B^{-} \to \pi ^{-} \bar{K}_{1}^{0} (1400)$ & $<3.9\times 10^{-5} $ \\ \hline 
$\bar{B}^{0} \to D_{s}^{-} a_{1}^{+} $ & $<2.1\times 10^{-3} $ \\ \hline 
$\bar{B}^{0} \to \pi ^{+} K_{1}^{-} (1270)$ & $<3.0\times 10^{-5} $ \\ \hline 
$\bar{B}^{0} \to \pi ^{+} K_{1}^{-} (1400)$ & $<2.7\times 10^{-5} $ \\ \hline 
$\bar{B}^{0} \to \pi ^{0} a_{1}^{0} $. & $<1.1\times 10^{-3} $ \\ \hline 
\end{tabular}

\end{table}

\clearpage

 \begin{table}
 
\centering
\caption{ The parameter$\beta $ for \textit{s}-wave and \textit{p}-wave mesons in the ISGW II model}
\label{t1}
\begin{tabular}{|c|c|c|c|c|c|c|c|c|} \hline 
\textbf{Quark content } & \boldmath$u\bar{d}$  & \boldmath$u\bar{s}$  & \boldmath$s\bar{s}$  & \boldmath$c\bar{u}$  &\boldmath $c\bar{s}$  & \boldmath$u\bar{b}$  &\boldmath $s\bar{b}$  & \boldmath$c\bar{c}$ \\ \hline 
\boldmath$\beta _{s} $(GeV)  & 0.41  & 0.44  & 0.53 & 0.45  & 0.56 & 0.43  & 0.54 & 0.88 \\ \hline 
\boldmath$\beta _{p} $(GeV)  & 0.28  & 0.30  & 0.33 & 0.33  & 0.38 & 0.35  & 0.41 & 0.52 \\ \hline 
\end{tabular}
\end{table}

 \begin{table}

\centering
\caption{ Form factors of $B(0^{-} )\to A(1^{+} )$ transition at $q^2$ =$t_m$  in the ISGW II quark model}
 \label{t2}
 
\begin{tabular}{|c|c|c|c|} \hline 
\textbf{Transition} & \textbf{\textit{l}} & \bm{$c_+$} & \bm{$c_-$} \\ \hline 
$B\to a_{1} $  & -2.38 & -0.032 & -0.0091 \\ \hline 
$B\to f_{1} $  & -2.38 & -0.032 & -0.0090 \\ \hline 
$B\to K_{1}^{} $ & -1.62 & -0.035 & -0.0074 \\ \hline 
$B\to D_{1} $ & -0.55 & -0.050 & -0.0041 \\ \hline 
\multicolumn{4}{|l|}{\textbf{HQS Constraints}} \\ \hline 
$B\to D_{1}^{1/2} $  & -0.014 & -0.090 & 0.094 \\ \hline 
$B\to D_{1}^{3/2} $ & -0.96 & -1.35 & 0.078 \\ \hline 
\end{tabular}

\end{table}

 \begin{table}
 
\centering
\caption{ Form factors of $B(0^{-} )\to A'(1^{+} )$ transition at $q^2$ =$t_m$ in the ISGW II quark model}
\label{t3}

\begin{tabular}{|c|c|c|c|} \hline 
\textbf{Transition} & \textbf{\textit{r}} & \boldmath$s_+$ & \boldmath$s_-$ \\ \hline 
$B\to b_{1} $  & 1.945 & 0.126 & -0.094 \\ \hline 
$B\to h_{1} $  & 1.908 & 0.128 & -0.096 \\ \hline 
$B\to \underline{K}_{1}^{} $ & 1.423 & 0.125 & -0.085 \\ \hline 
$B\to \underline{D}_{1} $ & 0.796 & 0.108 & -0.043 \\ \hline 
\end{tabular}

\end{table}

 \begin{table}

\centering
\caption{Form factors of $B(0^{-} )\to P(0^{-} )$ transition}
 \label{t4}

\begin{tabular}{|c|c|} \hline 
\textbf{Transition} & \boldmath $F_{0}^{BP} (0)$  \\ \hline 
$B\to \pi $ & 0.39 \\ \hline 
$B\to \eta $ & 0.37 \\ \hline 
$B\to K$ & 0.42 \\ \hline 
$B\to D$ & 0.70 \\ \hline 
\end{tabular}
\end{table}

\clearpage

 \begin{table}
 
\centering
\caption{ Decay Amplitudes ($\times \frac{G_{F} }{\sqrt{2} } V_{cb} V_{ud}^{*}$) for $B\to PA$ decays in CKM-favored mode involving $b\to c$ transition}

\label{t5}

\begin{tabular}{|c|c|} \hline 
\textbf{Decays} & \textbf{Amplitudes} \\ \hline 
\multicolumn{2}{|l|}{$\Delta b=1,  \Delta C=1,\Delta S=0 $} \\ \hline 
$B^{-} \to \pi ^{-} D_{1}^{0} $ & $\begin{array}{l} {                              a_{1} f_{\pi } (\sin \theta _{2} F^{B\to D_{1A} } (m_{\pi }^{2} )+\cos \theta _{2} F^{B\to D_{1A'} } (m_{\pi }^{2} ))} \\ {      +2m_{D_{1} } a_{2} (f_{D_{1A} } \sin \theta _{2} F^{B\to \pi } (m_{D_{1A} }^{2} )+f_{D_{1A'} } \cos \theta _{2} F^{B\to \pi } (m_{D_{1A'} }^{2} ))} \end{array}$ \\ \hline 
$B^{-} \to \pi ^{-} \underline{D}_{1}^{0} $  & $\begin{array}{l} {                                a_{1} f_{\pi } (\cos \theta _{2} F^{B\to \underline{D}_{1A} } (m_{\pi }^{2} )-\sin \theta _{2} F^{B\to \underline{D}_{1A'} } (m_{\pi }^{2} ))} \\ {      +2m_{\underline{D}_{1} } a_{2} (f_{\underline{D}_{1A} } \cos \theta _{2} F^{B\to \pi } (m_{\underline{D}_{1A} }^{2} )-f_{\underline{D}_{1A'} } \sin \theta _{2} F^{B\to \pi } (m_{\underline{D}_{1A'} }^{2} ))} \end{array}$ \\ \hline 
$B^{-} \to D^{0} a_{1}^{-} $ & $a_{2} f_{D} F^{B\to a_{1} } (m_{D}^{2} )+2a_{1} m_{a_{1} } f_{a_{1} } F^{B\to D} (m_{a_{1} }^{2} )$ \\ \hline 
$B^{-} \to D^{0} b_{1}^{-} $ & $a_{2} f_{D} F^{B\to b_{1} } (m_{D}^{2} )+2a_{1} m_{b_{1} } f_{b_{1} } F^{B\to D} (m_{b_{1} }^{2} )$ \\ \hline 
$\bar{B}^{0} \to \pi ^{0} D_{1}^{0} $ & $\sqrt{2} m_{D_{1} } a_{2} (-f_{D_{1A} } \sin \theta _{2} F^{\bar{B}\to \pi } (m_{D_{1} }^{2} )-f_{D_{1A'} } \cos \theta _{2} F^{\bar{B}\to \pi } (m_{D_{1} }^{2} ))$ \\ \hline 
$\bar{B}^{0} \to \pi ^{0} \underline{D}_{1}^{0} $ & $\sqrt{2} m_{D_{1} } a_{2} (-f_{D_{1A} } \cos \theta _{2} F^{\bar{B}\to \pi } (m_{D_{1} }^{2} )+f_{D_{1A'} } \sin \theta _{2} F^{\bar{B}\to \pi } (m_{D_{1} }^{2} ))$ \\ \hline 
$\bar{B}^{0} \to \pi ^{-} D_{1}^{+} $ & $a_{1} f_{\pi } (\sin \theta _{2} F^{\bar{B}\to D_{1A} } (m_{\pi }^{2} )+\cos \theta _{2} F^{\bar{B}\to D_{1A'} } (m_{\pi }^{2} ))$ \\ \hline 
$\bar{B}^{0} \to \pi ^{-} \underline{D}_{1}^{+} $ & $a_{1} f_{\pi } (\cos \theta _{2} F^{\bar{B}\to D_{1A} } (m_{\pi }^{2} )-\sin \theta _{2} F^{\bar{B}\to D_{1A'} } (m_{\pi }^{2} ))$ \\ \hline 
$\bar{B}^{0} \to \eta D_{1}^{0} $ & $\sqrt{2} m_{D_{1} } a_{2} (f_{D_{1A} } \sin \theta _{2} \sin \varphi _{P} F^{\bar{B}\to \eta } (m_{D_{1} }^{2} )+f_{D_{1A'} } \cos \theta _{2} \sin \varphi _{P} F^{\bar{B}\to \eta } (m_{D_{1} }^{2} ))$ \\ \hline 
$\bar{B}^{0} \to \eta \underline{D}_{1}^{0} $ & $\sqrt{2} m_{\underline{D}_{1} } a_{2} (f_{D_{1A} } \cos \theta _{2} \sin \varphi _{P} F^{\bar{B}\to \eta } (m_{\underline{D}_{1} }^{2} )-f_{D_{1A'} } \sin \theta _{2} \sin \varphi _{P} F^{\bar{B}\to \eta } (m_{\underline{D}_{1} }^{2} ))$ \\ \hline 
$\bar{B}^{0} \to \eta 'D_{1}^{0} $ & $\sqrt{2} m_{D_{1} } a_{2} (f_{D_{1A} } \sin \theta _{2} \cos \varphi _{P} F^{\bar{B}\to \eta '} (m_{D_{1} }^{2} )+f_{D_{1A'} } \cos \theta _{2} \cos \varphi _{P} F^{\bar{B}\to \eta '} (m_{D_{1} }^{2} ))$ \\ \hline 
$\bar{B}^{0} \to \eta '\underline{D}_{1}^{0} $ & $\sqrt{2} m_{\underline{D}_{1} } a_{2} (f_{D_{1A} } \cos \theta _{2} \cos \varphi _{P} F^{\bar{B}\to \eta '} (m_{\underline{D}_{1} }^{2} )-f_{D_{1A'} } \sin \theta _{2} \cos \varphi _{P} F^{\bar{B}\to \eta '} (m_{\underline{D}_{1} }^{2} ))$ \\ \hline 
$\bar{B}^{0} \to D^{+} a_{1}^{-} $ & $2a_{1} m_{a_{1} } f_{a_{1} } F^{\bar{B}\to D} (m_{a_{1} }^{2} )$ \\ \hline 
$\bar{B}^{0} \to D^{+} b_{1}^{-} $ & $2a_{1} m_{b_{1} } f_{b_{1} } F^{\bar{B}\to D} (m_{b_{1} }^{2} )$ \\ \hline 
$\bar{B}^{0} \to D^{0} a_{1}^{0} $  & $-(1/\sqrt{2} )a_{2} f_{D} F^{\bar{B}\to a_{1} } (m_{D}^{2} )$ \\ \hline 
$\bar{B}^{0} \to D^{0} f_{1} $  & $(1/\sqrt{2} )a_{2} f_{D} \cos \varphi _{A} F^{\bar{B}\to f_{1} } (m_{D}^{2} )$ \\ \hline 
$\bar{B}^{0} \to D^{0} b_{1}^{0} $  & $-(1/\sqrt{2} )a_{2} f_{D} F^{\bar{B}\to b_{1} } (m_{D}^{2} )$ \\ \hline 
$\bar{B}^{0} \to D^{0} h_{1} $  & $(1/\sqrt{2} )a_{2} f_{D} \cos \varphi _{A'} F^{\bar{B}\to h_{1} } (m_{D}^{2} )$ \\ \hline 
\end{tabular}

\end{table}

  \clearpage

 \begin{table}
 
\centering
\caption{ Decay Amplitudes ($\times \frac{G_{F} }{\sqrt{2} } V_{cb} V_{cs}^{*}$) for $B\to PA$ decays in CKM-favored mode involving $b\to c$ transition}
\label{t6}

\begin{tabular}{|c|c|} \hline 
\textbf{Decays} & \textbf{Amplitudes} \\ \hline 
\multicolumn{2}{|l|}{$\Delta b=1,  \Delta C=0,\Delta S=-1  $} \\ \hline 
$B^{-} \to K^{-} \chi _{c1} $ & $2m_{\chi _{c1} } a_{2} f_{\chi _{c1} } F_{}^{B\to K} (m_{\chi _{c1} }^{2} )$ \\ \hline 
$B^{-} \to D^{0} D_{s1}^{-} $ & $2a_{1} m_{D_{s1} } (f_{D_{s1A} } \sin \theta _{3} F^{B\to D} (m_{D_{s1} }^{2} )+f_{D_{s1A'} } \cos \theta _{3} F^{B\to D} (m_{D_{s1} }^{2} ))$ \\ \hline 
$B^{-} \to D^{0} \underline{D}_{s1}^{-} $ & $2a_{1} m_{\underline{D}_{s1} } (f_{D_{s1A} } \cos \theta _{3} F^{B\to D} (m_{\underline{D}_{s1} }^{2} )-f_{D_{s1A'} } \sin \theta _{3} F^{B\to D} (m_{\underline{D}_{s1} }^{2} ))$ \\ \hline 
$B^{-} \to D_{s}^{-} D_{1}^{0} $  & $a_{1} f_{D_{s} } (\sin \theta _{2} F^{B\to D_{1A} } (m_{D_{s} }^{2} )+\cos \theta _{2} F^{B\to D_{1A'} } (m_{D_{s} }^{2} ))$ \\ \hline 
$B^{-} \to D_{s}^{-} \underline{D}_{1}^{0} $  & $a_{1} f_{D_{s} } (\cos \theta _{2} F^{B\to D_{1A} } (m_{D_{s} }^{2} )-\sin \theta _{2} F^{B\to D_{1A'} } (m_{D_{s} }^{2} ))$ \\ \hline 
$B^{-} \to \eta _{c} K_{1}^{-} $ & $a_{2} f_{\eta _{c} } (\sin \theta _{1} F^{B\to K_{1A} } (m_{\eta _{c} }^{2} )+\cos \theta _{1} F^{B\to K_{1A'} } (m_{\eta _{c} }^{2} ))$ \\ \hline 
$B^{-} \to \eta _{c} \underline{K}_{1}^{-} $ & $a_{2} f_{\eta _{c} } (\cos \theta _{1} F^{B\to K_{1A} } (m_{\eta _{c} }^{2} )-\sin \theta _{1} F^{B\to K_{1A'} } (m_{\eta _{c} }^{2} ))$ \\ \hline 
$\bar{B}^{0} \to \bar{K}^{0} \chi _{c1} $ & $2m_{\chi _{c1} } a_{2} f_{\chi _{c1} } F_{}^{\bar{B}\to K} (m_{\chi _{c1} }^{2} )$ \\ \hline 
$\bar{B}^{0} \to D^{+} D_{s1}^{-} $ & $2a_{1} m_{D_{s1} } (f_{D_{s1A} } \sin \theta _{3} F^{\bar{B}\to D} (m_{D_{s1} }^{2} )+f_{D_{s1A'} } \cos \theta _{3} F^{\bar{B}\to D} (m_{D_{s1} }^{2} ))$ \\ \hline 
$\bar{B}^{0} \to D^{+} \underline{D}_{s1}^{-} $ & $2a_{1} m_{\underline{D}_{s1} } (f_{D_{s1A} } \sin \theta _{3} F^{\bar{B}\to D} (m_{\underline{D}_{s1} }^{2} )+f_{D_{s1A'} } \cos \theta _{3} F^{\bar{B}\to D} (m_{\underline{D}_{s1} }^{2} ))$ \\ \hline 
$\bar{B}^{0} \to D_{s}^{-} D_{1}^{+} $  & $a_{1} f_{D_{s} } (\sin \theta _{2} F^{B\to D_{1A} } (m_{D_{s} }^{2} )+\cos \theta _{2} F^{B\to D_{1A'} } (m_{D_{s} }^{2} ))$ \\ \hline 
$\bar{B}^{0} \to D_{s}^{-} \underline{D}_{1}^{+} $  & $a_{1} f_{D_{s} } (\cos \theta _{2} F^{B\to D_{1A} } (m_{D_{s} }^{2} )-\sin \theta _{2} F^{B\to D_{1A'} } (m_{D_{s} }^{2} ))$ \\ \hline 
$\bar{B}^{0} \to \eta _{c} \bar{K}_{1}^{0} $  & $a_{2} f_{\eta _{c} } (\sin \theta _{1} F^{\bar{B}\to K_{1A} } (m_{\eta _{c} }^{2} )+\cos \theta _{1} F^{\bar{B}\to K_{1A'} } (m_{\eta _{c} }^{2} ))$ \\ \hline 
$\bar{B}^{0} \to \eta _{c} \underline{\bar{K}}_{1}^{0} $  & $a_{2} f_{\eta _{c} } (\cos \theta _{1} F^{\bar{B}\to K_{1A} } (m_{\eta _{c} }^{2} )-\sin \theta _{1} F^{\bar{B}\to K_{1A'} } (m_{\eta _{c} }^{2} ))$ \\ \hline 
\end{tabular}

\end{table}
\clearpage

 \begin{table}
 
\centering
\caption{ Decay Amplitudes ($\times \frac{G_{F} }{\sqrt{2} } V_{cb} V_{us}^{*}$) for $B\to PA$ decays in CKM-suppressed mode involving $b\to c$ transition}

 \label{t7} 

\begin{tabular}{|c|c|} \hline 
\textbf{Decays} & \textbf{Amplitudes} \\ \hline 
\multicolumn{2}{|l|}{$\Delta b=1,  \Delta C=1,\Delta S=-1 $ } \\ \hline 
$B^{-} \to K^{-} D_{1}^{0} $ & $\begin{array}{l} {                            a_{1} f_{K} (\sin \theta _{2} F^{B\to D_{1A} } (m_{K}^{2} )+\cos \theta _{2} F^{B\to D_{1A'} } (m_{K}^{2} ))} \\ {      +2m_{D_{1} } a_{2} (f_{D_{1A} } \sin \theta _{2} F^{B\to K} (m_{D_{1} }^{2} )+f_{D_{1A'} } \cos \theta _{2} F^{B\to K} (m_{D_{1} }^{2} ))} \end{array}$ \\ \hline 
$B^{-} \to K^{-} \underline{D}_{1}^{0} $  & $\begin{array}{l} {                            a_{1} f_{K} (\cos \theta _{2} F^{B\to D_{1A} } (m_{K}^{2} )-\sin \theta _{2} F^{B\to D_{1A'} } (m_{K}^{2} ))} \\ {      +2m_{\underline{D}_{1} } a_{2} (f_{D_{1A} } \cos \theta _{2} F^{B\to K} (m_{\underline{D}_{1} }^{2} )-f_{D_{1A'} } \sin \theta _{2} F^{B\to K} (m_{\underline{D}_{1} }^{2} ))} \end{array}$ \\ \hline 
$B^{-} \to D^{0} K_{1}^{-} $ & $\begin{array}{l} {                            a_{2} f_{D} (\sin \theta _{1} F^{B\to K_{1A} } (m_{D}^{2} )+\cos \theta _{1} F^{B\to K_{1A'} } (m_{D}^{2} ))} \\ {      +2m_{K_{1} } a_{1} (f_{K_{1A} } \sin \theta _{1} F^{B\to D} (m_{K_{1} }^{2} )+f_{K_{1A'} } \cos \theta _{1} F^{B\to D} (m_{K_{1} }^{2} ))} \end{array}$ \\ \hline 
$B^{-} \to D^{0} \underline{K}_{1}^{-} $ & $\begin{array}{l} {                          a_{2} f_{D} (\cos \theta _{1} F^{B\to K_{1A} } (m_{D}^{2} )-\sin \theta _{1} F^{B\to K_{1A'} } (m_{D}^{2} ))} \\ {      +2m_{K_{1} } a_{1} (f_{K_{1A} } \cos \theta _{1} F^{B\to D} (m_{K_{1} }^{2} )-f_{K_{1A'} } \sin \theta _{1} F^{B\to D} (m_{K_{1} }^{2} ))} \end{array}$ \\ \hline 
$\bar{B}^{0} \to \bar{K}^{0} D_{1}^{0} $ & $2m_{D_{1} } a_{2} (f_{D_{1A} } \sin \theta _{2} F^{\bar{B}\to K} (m_{D_{1} }^{2} )+f_{D_{1A'} } \cos \theta _{2} F^{\bar{B}\to K} (m_{D_{1} }^{2} ))$ \\ \hline 
$\bar{B}^{0} \to \bar{K}^{0} \underline{D}_{1}^{0} $ & $2m_{\underline{D}_{1} } a_{2} (f_{D_{1A} } \cos \theta _{2} F^{\bar{B}\to K} (m_{\underline{D}_{1} }^{2} )-f_{D_{1A'} } \sin \theta _{2} F^{\bar{B}\to K} (m_{\underline{D}_{1} }^{2} ))$ \\ \hline 
$\bar{B}^{0} \to K^{-} D_{1}^{+} $ & $a_{1} f_{K} (\sin \theta _{2} F^{\bar{B}\to D_{1A} } (m_{K}^{2} )+\cos \theta _{2} F^{\bar{B}\to D_{1A'} } (m_{K}^{2} ))$ \\ \hline 
$\bar{B}^{0} \to K^{-} \underline{D}_{1}^{+} $ & $a_{1} f_{K} (\cos \theta _{2} F^{\bar{B}\to D_{1A} } (m_{K}^{2} )-\sin \theta _{2} F^{\bar{B}\to D_{1A'} } (m_{K}^{2} ))$ \\ \hline 
$\bar{B}^{0} \to D^{+} K_{1}^{-} $ & $2m_{K_{1} } a_{1} (f_{K_{1A} } \sin \theta _{1} F^{\bar{B}\to D} (m_{K_{1} }^{2} )+f_{K_{1A'} } \cos \theta _{1} F^{\bar{B}\to D} (m_{K_{1} }^{2} ))$ \\ \hline 
$\bar{B}^{0} \to D^{+} \underline{K}_{1}^{-} $ & $2m_{\underline{K}_{1} } a_{1} (f_{K_{1A} } \cos \theta _{1} F^{\bar{B}\to D} (m_{\underline{K}_{1} }^{2} )-f_{K_{1A'} } \sin \theta _{1} F^{\bar{B}\to D} (m_{\underline{K}_{1} }^{2} ))$ \\ \hline 
$\bar{B}^{0} \to D^{0} \bar{K}_{1}^{0} $ & $a_{2} f_{D} (\sin \theta _{1} F^{\bar{B}\to K_{1A} } (m_{D}^{2} )+\cos \theta _{1} F^{\bar{B}\to K_{1A'} } (m_{D}^{2} ))$ \\ \hline 
$\bar{B}^{0} \to D^{0} \bar{\underline{K}}_{1}^{0} $ & $a_{2} f_{D} (\cos \theta _{1} F^{\bar{B}\to K_{1A} } (m_{D}^{2} )-\sin \theta _{1} F^{\bar{B}\to K_{1A'} } (m_{D}^{2} ))$ \\ \hline 
\end{tabular}

 \end{table}
\clearpage

 \begin{table}
 
\centering
\caption{Decay Amplitudes ($\times \frac{G_{F} }{\sqrt{2} } V_{cb} V_{cd}^{*}$) for $B\to PA$ decays in CKM-suppressed mode involving $b\to c$ transition}
\label{t8}

\begin{tabular}{|c|c|} \hline 
\textbf{Decays} & \textbf{Amplitudes} \\ \hline 
\multicolumn{2}{|l|}{$\Delta b=1,  \Delta C=0,\Delta S=0 $} \\ \hline 
$B^{-} \to \pi ^{-} \chi _{c1} $ &  \textbf{-}$2m_{\chi _{c1} } a_{2} f_{\chi _{c1} } F_{}^{B\to \pi } (m_{\chi _{c1} }^{2} )$ \\ \hline 
$B^{-} \to D^{0} D_{1}^{-} $ & - $2m_{D_{1} } a_{1} (f_{D_{1A} } \sin \varphi _{2} F^{B\to D} (m_{D_{1} }^{2} )+f_{D_{1A'} } \cos \varphi _{2} F^{B\to D} (m_{D_{1} }^{2} ))$ \\ \hline 
$B^{-} \to D^{0} \underline{D}_{1}^{-} $ & - $2m_{\underline{D}_{1} } a_{1} (f_{D_{1A} } \cos \theta _{2} F^{B\to D} (m_{\underline{D}_{1} }^{2} )-f_{D_{1A'} } \sin \theta _{2} F^{B\to D} (m_{\underline{D}_{1} }^{2} ))$ \\ \hline 
$B^{-} \to D^{-} D_{1}^{0} $ & - $a_{1} f_{D} (\sin \theta _{2} F^{B\to D_{1A} } (m_{D}^{2} )+\cos \theta _{2} F^{B\to D_{1A'} } (m_{D}^{2} ))$ \\ \hline 
$B^{-} \to D^{-} \underline{D}_{1}^{0} $ & - $a_{1} f_{D} (\cos \theta _{2} F^{B\to D_{1A} } (m_{D}^{2} )-\sin \theta _{2} F^{B\to D_{1A'} } (m_{D}^{2} ))$ \\ \hline 
$B^{-} \to \eta _{c} a_{1}^{-} $ & - $a_{2} f_{\eta _{c} } F^{B\to a_{1} } (m_{\eta _{c} }^{2} )$ \\ \hline 
$B^{-} \to \eta _{c} b_{1}^{-} $ & - $a_{2} f_{\eta _{c} } F^{B\to b_{1} } (m_{\eta _{c} }^{2} )$ \\ \hline 
$\bar{B}^{0} \to \pi ^{0} \chi _{c1} $ &  $\sqrt{2} m_{\chi _{c1} } a_{2} f_{\chi _{c1} } F_{}^{\bar{B}\to \pi } (m_{\chi _{c1} }^{2} )$ \\ \hline 
$\bar{B}^{0} \to \eta \chi _{c1} $ & \textbf{                         - }$\sqrt{2} m_{\chi _{c1} } a_{2} f_{\chi _{c1} } \sin \varphi _{P} F_{}^{\bar{B}\to \eta } (m_{\chi _{c1} }^{2} )$ \\ \hline 
$\bar{B}^{0} \to \eta '\chi _{c1} $  &  -$\sqrt{2} m_{\chi _{c1} } a_{2} f_{\chi _{c1} } \cos \varphi _{P} F_{}^{\bar{B}\to \eta '} (m_{\chi _{c1} }^{2} )$ \\ \hline 
$\bar{B}^{0} \to D^{+} D_{1}^{-} $ & -$2m_{D_{1} } a_{1} (f_{D_{1A} } \sin \theta _{2} F^{\bar{B}\to D} (m_{D_{1} }^{2} )+f_{D_{1A'} } \cos \theta _{2} F^{\bar{B}\to D} (m_{D_{1} }^{2} ))$ \\ \hline 
$\bar{B}^{0} \to D^{+} \underline{D}_{1}^{-} $ & -$2m_{D_{1} } a_{1} (f_{D_{1A} } \cos \theta _{2} F^{\bar{B}\to D} (m_{D_{1} }^{2} )-f_{D_{1A'} } \sin \theta _{2} F^{\bar{B}\to D} (m_{D_{1} }^{2} ))$ \\ \hline 
$\bar{B}^{0} \to D^{-} D_{1}^{+} $ & -$a_{1} f_{D} (\sin \theta _{2} F^{\bar{B}\to D_{1A} } (m_{D}^{2} )+\cos \theta _{2} F^{\bar{B}\to D_{1A'} } (m_{D}^{2} ))$ \\ \hline 
$\bar{B}^{0} \to D^{-} \underline{D}_{1}^{+} $ & -$a_{1} f_{D} (\cos \theta _{2} F^{B\to D_{1A} } (m_{D}^{2} )-\sin \theta _{2} F^{B\to D_{1A'} } (m_{D}^{2} ))$ \\ \hline 
$\bar{B}^{0} \to \eta _{c} a_{1}^{0} $  & $(1/\sqrt{2} )a_{2} f_{\eta _{c} } F^{\bar{B}\to a_{1} } (m_{\eta _{c} }^{2} )$ \\ \hline 
$\bar{B}^{0} \to \eta _{c} f_{1} $  & -$(1/\sqrt{2} )a_{2} f_{\eta _{c} } \cos \varphi _{A} F^{\bar{B}\to f_{1} } (m_{\eta _{c} }^{2} )$ \\ \hline 
$\bar{B}^{0} \to \eta _{c} b_{1}^{0} $ & $(1/\sqrt{2} )a_{2} f_{\eta _{c} } F^{\bar{B}\to b_{1} } (m_{\eta _{c} }^{2} )$ \\ \hline 
$\bar{B}^{0} \to \eta _{c} h_{1} $ & -$(1/\sqrt{2} )a_{2} f_{\eta _{c} } \cos \varphi _{A'} F^{\bar{B}\to h_{1} } (m_{\eta _{c} }^{2} )$ \\ \hline 
\end{tabular}

 \end{table}
\clearpage

 \begin{table}

\centering
\caption{Decay Amplitudes ($\times \frac{G_{F} }{\sqrt{2} } V_{ub} V_{cs}^{*} $) for $B\to PA$ decays involving $b\to u$ transition}

   \label{t9}

\begin{tabular}{|c|c|} \hline 
\textbf{Decays} & \textbf{Amplitudes} \\ \hline 
\multicolumn{2}{|l|}{$\Delta b=1,  \Delta C=-1,\Delta S=-1$} \\ \hline 
$B^{-} \to \pi ^{0} D_{s1}^{-} $ & $\sqrt{2} m_{D_{s1} } a_{1} (f_{D_{1sA} } \sin \theta _{3} F_{}^{B\to \pi } (m_{D_{s1} }^{2} )+f_{D_{s1A'} } \cos \theta _{3} F_{}^{B\to \pi } (m_{D_{s1} }^{2} ))$ \\ \hline 
$B^{-} \to \pi ^{0} \underline{D}_{s1}^{-} $  & $\sqrt{2} m_{\underline{D}_{s1} } a_{1} (f_{D_{1sA} } \cos \theta _{3} F_{}^{B\to \pi } (m_{\underline{D}_{s1} }^{2} )-f_{D_{s1A'} } \sin \theta _{3} F_{}^{B\to \pi } (m_{\underline{D}_{s1} }^{2} ))$ \\ \hline 
$B^{-} \to \eta D_{s1}^{-} $ & $\sqrt{2} m_{D_{s1} } a_{1} (f_{D_{1sA} } \sin \varphi _{P} \sin \theta _{3} F_{}^{B\to \eta } (m_{D_{s1} }^{2} )+f_{D_{s1A'} } \cos \varphi _{P} \cos \theta _{3} F_{}^{B\to \eta } (m_{D_{s1} }^{2} ))$ \\ \hline 
$B^{-} \to \eta \underline{D}_{s1}^{-} $ & $\sqrt{2} m_{\underline{D}_{s1} } a_{1} (f_{D_{1sA} } \sin \varphi _{P} \cos \theta _{3} F_{}^{B\to \eta } (m_{\underline{D}_{s1} }^{2} )-f_{D_{s1A'} } \cos \varphi _{P} \sin \theta _{3} F_{}^{B\to \eta } (m_{\underline{D}_{s1} }^{2} ))$ \\ \hline 
$B^{-} \to K^{-} \bar{D}_{1}^{0} $ & $2m_{D_{1} } a_{2} (f_{D_{1A} } \sin \theta _{2} F_{}^{B\to K} (m_{D_{1} }^{2} )+f_{D_{1A'} } \cos \theta _{2} F_{}^{B\to K} (m_{D_{1} }^{2} ))$ \\ \hline 
$B^{-} \to K^{-} \underline{\bar{D}}_{1}^{0} $ & $2m_{\underline{D}_{1} } a_{2} (f_{D_{1A} } \cos \theta _{2} F_{}^{B\to K} (m_{\underline{D}_{1} }^{2} )-f_{D_{1A'} } \sin \theta _{2} F_{}^{B\to K} (m_{\underline{D}_{1} }^{2} ))$ \\ \hline 
$B^{-} \to \eta 'D_{s1}^{-} $ & $\sqrt{2} m_{D_{s1} } a_{1} (f_{D_{1sA} } \cos \varphi _{P} \sin \theta _{3} F_{}^{B\to \eta '} (m_{D_{s1} }^{2} )+f_{D_{s1A'} } \cos \varphi _{P} \cos \theta _{3} F_{}^{B\to \eta '} (m_{D_{s1} }^{2} ))$ \\ \hline 
$B^{-} \to \eta '\underline{D}_{s1}^{-} $ & $\sqrt{2} m_{\underline{D}_{s1} } a_{1} (f_{D_{1sA} } \cos \varphi _{P} \cos \theta _{3} F_{}^{B\to \eta '} (m_{\underline{D}_{s1} }^{2} )-f_{D_{s1A'} } \cos \varphi _{P} \sin \theta _{3} F_{}^{B\to \eta '} (m_{\underline{D}_{s1} }^{2} ))$ \\ \hline 
$B^{-} \to \bar{D}^{0} K_{1}^{-} $ & $a_{2} f_{D} (\sin \theta _{1} F_{}^{B\to K_{1A} } (m_{D}^{2} )+\cos \theta _{1} F_{}^{B\to K_{1A'} } (m_{D}^{2} ))$ \\ \hline 
$B^{-} \to \bar{D}^{0} \underline{K}_{1}^{-} $ & $a_{2} f_{D} (\cos \theta _{1} F_{}^{B\to K_{1A} } (m_{D}^{2} )-\sin \theta _{1} F_{}^{B\to K_{1A'} } (m_{D}^{2} ))$ \\ \hline 
$B^{-} \to D_{s}^{-} a_{1}^{0} $ & $(1/\sqrt{2} )  a_{1} f_{D_{s} } F_{}^{B\to a_{1} } (m_{D_{s} }^{2} )$ \\ \hline 
$B^{-} \to D_{s}^{-} f_{1} $ & $(1/\sqrt{2} )a_{1} f_{D_{s} } \cos \varphi _{A} F_{}^{B\to f_{1} } (m_{D_{s} }^{2} )$ \\ \hline 
$B^{-} \to D_{s}^{-} b_{1}^{0} $ & $(1/\sqrt{2} )a_{1} f_{D_{s} } F_{}^{B\to b_{1} } (m_{D_{s} }^{2} )$ \\ \hline 
$B^{-} \to D_{s}^{-} h_{1} $ & $(1/\sqrt{2} )a_{1} f_{D_{s} } \cos \varphi _{A'} F_{}^{B\to h_{1} } (m_{D_{s} }^{2} )$ \\ \hline 
$\bar{B}^{0} \to \pi ^{+} D_{s1}^{-} $  & $2m_{D_{s1} } a_{1} (f_{D_{1sA} } \sin \theta _{3} F_{}^{\bar{B}\to \pi } (m_{D_{s1} }^{2} )+f_{D_{s1A'} } \cos \theta _{3} F_{}^{\bar{B}\to \pi } (m_{D_{s1} }^{2} ))$ \\ \hline 
$\bar{B}^{0} \to \pi ^{+} \underline{D}_{s1}^{-} $  & $2m_{\underline{D}_{s1} } a_{1} (f_{D_{1sA} } \cos \theta _{3} F_{}^{\bar{B}\to \pi } (m_{\underline{D}_{s1} }^{2} )-f_{D_{s1A'} } \sin \theta _{3} F_{}^{\bar{B}\to \pi } (m_{\underline{D}_{s1} }^{2} ))$ \\ \hline 
$\bar{B}^{0} \to \bar{K}^{0} \bar{D}_{1}^{0} $  & $2m_{D_{1} } a_{2} (f_{D_{1A} } \sin \theta _{2} F_{}^{\bar{B}\to K} (m_{D_{1} }^{2} )+f_{D_{1A'} } \cos \theta _{2} F_{}^{\bar{B}\to K} (m_{D_{1} }^{2} ))$ \\ \hline 
$\bar{B}^{0} \to \bar{K}^{0} \underline{\bar{D}}_{1}^{0} $  & $2m_{\underline{D}_{1} } a_{2} (f_{D_{1A} } \cos \theta _{2} F_{}^{\bar{B}\to K} (m_{\underline{D}_{1} }^{2} )-f_{D_{1A'} } \sin \theta _{2} F_{}^{\bar{B}\to K} (m_{\underline{D}_{1} }^{2} ))$ \\ \hline 
$\bar{B}^{0} \to \bar{D}^{0} \bar{K}_{1}^{0} $ & $a_{2} f_{D} (\sin \theta _{1} F_{}^{\bar{B}\to K_{1A} } (m_{D}^{2} )+\cos \theta _{1} F_{}^{\bar{B}\to K_{1A'} } (m_{D}^{2} ))$ \\ \hline 
$\bar{B}^{0} \to \bar{D}^{0} \underline{\bar{K}}_{1}^{0} $ & $a_{2} f_{D} (\cos \theta _{1} F_{}^{\bar{B}\to K_{1A} } (m_{D}^{2} )-\sin \theta _{1} F_{}^{\bar{B}\to K_{1A'} } (m_{D}^{2} ))$ \\ \hline 
$\bar{B}^{0} \to D_{s}^{-} a_{1}^{+} $ & $a_{1} f_{D_{s} } F_{}^{B\to a_{1} } (m_{D_{s} }^{2} )$ \\ \hline 
$\bar{B}^{0} \to D_{s}^{-} b_{1}^{+} $ & $a_{1} f_{D_{s} } F_{}^{B\to b_{1} } (m_{D_{s} }^{2} )$ \\ \hline 
\end{tabular}

\end{table}
\clearpage

 \begin{table}
 
\centering
\caption{Decay Amplitudes ($\times \frac{G_{F} }{\sqrt{2} } V_{ub} V_{cd}^{*} $) for $B\to PA$ decays involving $b\to u$  transition}
\label{t10}
\begin{tabular}{|c|c|} \hline 
\textbf{Decays} & \textbf{Amplitudes} \\ \hline 
\multicolumn{2}{|l|}{$\Delta b=1,  \Delta C=-1,\Delta S=0$} \\ \hline 
$B^{-} \to \pi ^{0} D_{1}^{-} $ & \textbf{-}$\sqrt{2} m_{D_{1} } a_{1} (f_{D_{1A} } \sin \theta _{2} F_{}^{B\to \pi } (m_{D_{1} }^{2} )+f_{D_{1A'} } \cos \theta _{2} F_{}^{B\to \pi } (m_{D_{1} }^{2} ))$ \\ \hline 
$B^{-} \to \pi ^{0} \underline{D}_{1}^{-} $ & \textbf{- }$\sqrt{2} m_{\underline{D}_{1} } a_{1} (f_{D_{1A} } \cos \theta _{2} F_{}^{B\to \pi } (m_{\underline{D}_{1} }^{2} )-f_{D_{1A'} } \sin \theta _{2} F_{}^{B\to \pi } (m_{\underline{D}_{1} }^{2} ))$ \\ \hline 
$B^{-} \to \pi ^{-} \bar{D}_{1}^{0} $  & \textbf{-}$2m_{D_{1} } a_{2} (f_{D_{1A} } \sin \theta _{2} F_{}^{B\to \pi } (m_{\underline{D}_{1} }^{2} )+f_{D_{1A'} } \cos \theta _{2} F_{}^{B\to \pi } (m_{\underline{D}_{1} }^{2} ))$ \\ \hline 
$B^{-} \to \pi ^{-} \underline{\bar{D}}_{1}^{0} $  & \textbf{-}$2m_{\underline{D}_{1} } a_{2} (f_{D_{1A} } \cos \theta _{2} F_{}^{B\to \pi } (m_{\underline{D}_{1} }^{2} )-f_{D_{1A'} } \sin \theta _{2} F_{}^{B\to \pi } (m_{\underline{D}_{1} }^{2} ))$ \\ \hline 
$B^{-} \to \eta D_{1}^{-} $ & \textbf{-}$\sqrt{2} m_{D_{1} } a_{1} (f_{D_{1A} } \sin \theta _{2} \sin \varphi _{P} F_{}^{B\to \eta } (m_{D_{1} }^{2} )+f_{D_{1A'} } \cos \theta _{2} \sin \varphi _{P} F_{}^{B\to \eta } (m_{D_{1} }^{2} ))$ \\ \hline 
$B^{-} \to \eta \underline{D}_{1}^{-} $  & \textbf{-}$\sqrt{2} m_{\underline{D}_{1} } a_{1} (f_{D_{1A} } \cos \theta _{2} \sin \varphi _{P} F_{}^{B\to \eta } (m_{\underline{D}_{1} }^{2} )-f_{D_{1A'} } \sin \theta _{2} \sin \varphi _{P} F_{}^{B\to \eta } (m_{\underline{D}_{1} }^{2} ))$ \\ \hline 
$B^{-} \to \eta 'D_{1}^{-} $ & \textbf{-}$\sqrt{2} m_{D_{1} } a_{1} (f_{D_{1A} } \cos \theta _{2} \sin \varphi _{P} F_{}^{B\to \eta '} (m_{D_{1} }^{2} )+f_{D_{1A'} } \cos \theta _{2} \cos \varphi _{P} F_{}^{B\to \eta '} (m_{D_{1} }^{2} ))$ \\ \hline 
$B^{-} \to \eta '\underline{D}_{1}^{-} $ & \textbf{-}$\sqrt{2} m_{\underline{D}_{1} } a_{1} (f_{D_{1A} } \cos \theta _{2} \cos \varphi _{P} F_{}^{B\to \eta '} (m_{\underline{D}_{1} }^{2} )-f_{D_{1B} } \sin \theta _{2} \cos \varphi _{P} F_{}^{B\to \eta '} (m_{\underline{D}_{1} }^{2} ))$ \\ \hline 
$B^{-} \to D^{-} a_{1}^{0} $  & $-(1/\sqrt{2} )a_{1} f_{D} F_{}^{B\to a_{1} } (m_{D}^{2} )$ \\ \hline 
$B^{-} \to D^{-} f_{1} $  & $-(1/\sqrt{2} )a_{1} f_{D} \cos \varphi _{A} F_{}^{B\to f_{1} } (m_{D}^{2} )$  \\ \hline 
$B^{-} \to D^{-} b_{1}^{0} $  & \textbf{-}$(1/\sqrt{2} )a_{1} f_{D} F_{}^{B\to b_{1} } (m_{D}^{2} )$ \\ \hline 
$B^{-} \to D^{-} h_{1} $ & \textbf{-}$(1/\sqrt{2} )a_{1} f_{D} \cos \varphi _{A'} F_{}^{B\to h_{1} } (m_{D}^{2} )$  \\ \hline 
$B^{-} \to \bar{D}^{0} a_{1}^{-} $ & $-a_{2} f_{D} F_{}^{B\to a_{1} } (m_{D}^{2} )$ \\ \hline 
$B^{-} \to \bar{D}^{0} b_{1}^{-} $ & $-a_{2} f_{D} F_{}^{B\to b_{1} } (m_{D}^{2} )$ \\ \hline 
$\bar{B}^{0} \to \pi ^{+} D_{1}^{-} $ & $-2m_{D_{1} } a_{1} (f_{D_{1A} } \sin \theta _{2} F_{}^{\bar{B}\to \pi } (m_{D_{1} }^{2} )+f_{D_{1A'} } \cos \theta _{2} F_{}^{\bar{B}\to \pi } (m_{D_{1} }^{2} ))$ \\ \hline 
$\bar{B}^{0} \to \pi ^{+} \underline{D}_{1}^{-} $ & $-2m_{\underline{D}_{1} } a_{1} (f_{D_{1A} } \cos \theta _{2} F_{}^{B\to \pi } (m_{D_{1} }^{2} )-f_{D_{1A'} } \sin \theta _{2} F_{}^{B\to \pi } (m_{\underline{D}_{1} }^{2} ))$ \\ \hline 
$\bar{B}^{0} \to \pi ^{0} \bar{D}_{1}^{0} $ & $-\sqrt{2} m_{D_{1} } a_{2} (-f_{D_{1A} } \sin \theta _{2} F_{}^{\bar{B}\to \pi } (m_{D_{1} }^{2} )-f_{D_{1A'} } \cos \theta _{2} F_{}^{\bar{B}\to \pi } (m_{D_{1} }^{2} ))$ \\ \hline 
$\bar{B}^{0} \to \pi ^{0} \underline{\bar{D}}_{1}^{0} $ & $-\sqrt{2} m_{\underline{D}_{1} } a_{2} (-f_{D_{1A} } \cos \theta _{2} F_{}^{\bar{B}\to \pi } (m_{D_{1} }^{2} )+f_{D_{1A'} } \sin \theta _{2} F_{}^{\bar{B}\to \pi } (m_{\underline{D}_{1} }^{2} ))$ \\ \hline 
$\bar{B}^{0} \to \eta \bar{D}_{1}^{0} $ & $-\sqrt{2} m_{D_{1} } a_{2} (f_{\underline{D}_{1A} } \sin \theta _{2} \sin \varphi _{P} F_{}^{\bar{B}\to \eta } (m_{D_{1} }^{2} )+f_{D_{1A'} } \cos \theta _{2} \sin \varphi _{P} F_{}^{\bar{B}\to \eta } (m_{D_{1} }^{2} ))$ \\ \hline 
$\bar{B}^{0} \to \eta \underline{\bar{D}}_{1}^{0} $ & $-\sqrt{2} m_{\underline{D}_{1} } a_{2} (f_{D_{1A} } \cos \theta _{2} \sin \varphi _{P} F_{}^{\bar{B}\to \eta } (m_{\underline{D}_{1} }^{2} )-f_{D_{1A'} } \sin \theta _{2} \sin \varphi _{P} F_{}^{\bar{B}\to \eta } (m_{\underline{D}_{1} }^{2} ))$ \\ \hline 
$\bar{B}^{0} \to \eta '\bar{D}_{1}^{0} $ & $-\sqrt{2} m_{D_{1} } a_{2} (f_{D_{1A} } \sin \theta _{2} \cos \varphi _{P} F_{}^{\bar{B}\to \eta '} (m_{D_{1} }^{2} )+f_{D_{1A'} } \cos \theta _{2} \cos \varphi _{P} F_{}^{\bar{B}\to \eta '} (m_{D_{1} }^{2} ))$\\ \hline 
$\bar{B}^{0} \to \eta '\underline{\bar{D}}_{1}^{0} $ & $-\sqrt{2} m_{\underline{D}_{1} } a_{2} (f_{D_{1A} } \cos \theta _{2} \cos \varphi _{P} F_{}^{\bar{B}\to \eta '} (m_{\underline{D}_{1} }^{2} )-f_{D_{1A'} } \sin \theta _{2} \cos \varphi _{P} F_{}^{\bar{B}\to \eta '} (m_{\underline{D}_{1} }^{2} ))$ \\ \hline 
$\bar{B}^{0} \to D^{-} a_{1}^{+} /D^{-} b_{1}^{+} $  & $-a_{1} f_{D} F_{}^{\bar{B}\to a_{1} /b_{1} } (m_{D}^{2} )$ \\ \hline 
$\bar{B}^{0} \to \bar{D}^{0} a_{1}^{0} $ & $(1/\sqrt{2} )a_{2} f_{D} F_{}^{\bar{B}\to a_{1} } (m_{D}^{2} )$ \\ \hline 
$\bar{B}^{0} \to \bar{D}^{0} f_{1} $ & $-(1/\sqrt{2} )a_{2} f_{D} \cos \varphi _{A} F_{}^{\bar{B}\to f_{1} } (m_{D}^{2} )$ \\ \hline 
$\bar{B}^{0} \to \bar{D}^{0} b_{1}^{0} $ & $-(1/\sqrt{2} )a_{2} f_{D} F_{}^{\bar{B}\to b_{1} } (m_{D}^{2} )$ \\ \hline 
$\bar{B}^{0} \to \bar{D}^{0} h_{1} $ & $-(1/\sqrt{2} )a_{2} f_{D} \cos \varphi _{A'} F_{}^{\bar{B}\to h_{1} } (m_{D}^{2} )$ \\ \hline 
\end{tabular}

\end{table}
\clearpage

 \begin{table}
 
\centering
\caption{Branching ratios for $B\to PA$ decays in CKM-favored mode involving $b\to c$ transition. Numbers in [ ] are experimental values.}
\label{t11}

\begin{tabular}{|l|l|} \hline 
\textbf{Decays} & \textbf{Branching ratios} \\ \hline 
\multicolumn{2}{|l|}{$\Delta b=1,  \Delta C=1,\Delta S=0$ } \\ \hline 
$B^{-} \to \pi ^{-} D_{1}^{0} $  & 8.3$\times10^{-4}$[(7.5 $\pm$ 1.7) $\times 10^{-4}$\cite{32}] \\ \hline 
$B^{-} \to \pi ^{-} \underline{D}_{1}^{0} $  & 1.4$\times10^{-3}$[$(1.5\pm 0.6)\times 10^{-3} $] \\ \hline 
$B^{-} \to D^{0} a_{1}^{-} $  & 5.5$\times10^{-3}$[$(4.0\pm 4.0)\times 10^{-3} $] \\ \hline 
$B^{-} \to D^{0} b_{1}^{-} $  & 6.5$\times10^{-4}$ \\ \hline 
$\bar{B}^{0} \to \pi ^{0} D_{1}^{0} $  & 4.6$\times10^{-7}$ \\ \hline 
$\bar{B}^{0} \to \pi ^{0} \underline{D}_{1}^{0} $  & 5.9$\times10^{-5}$ \\ \hline 
$\bar{B}^{0} \to \pi ^{-} D_{1}^{+} $  & 8.3$\times10^{-4}$ \\ \hline 
$\bar{B}^{0} \to \pi ^{-} \underline{D}_{1}^{+} $  & 2.2$\times10^{-3}$ \\ \hline 
$\bar{B}^{0} \to \eta D_{1}^{0} $  & 2.4$\times10^{-7}$ \\ \hline 
$\bar{B}^{0} \to \eta \underline{D}_{1}^{0} $  & 3.0$\times10^{-5}$ \\ \hline 
$\bar{B}^{0} \to \eta 'D_{1}^{0} $  & 1.1$\times10^{-7}$ \\ \hline 
$\bar{B}^{0} \to \eta '\underline{D}_{1}^{0} $  & 1.4$	\times10^{-5}$ \\ \hline 
$\bar{B}^{0} \to D^{+} a_{1}^{-} $  & 1.1$\times10^{-2}$ [$(0.60\pm 0.33)\times 10^{-2} $] \\ \hline 
$\bar{B}^{0} \to D^{+} b_{1}^{-} $  & 9.9$	\times10^{-8}$ \\ \hline 
$\bar{B}^{0} \to D^{0} a_{1}^{0} $  & 5.7$	\times10^{-4}$\\ \hline 
$\bar{B}^{0} \to D^{0} f_{1} $  & 5.2$	\times10^{-4}$\\ \hline 
$\bar{B}^{0} \to D^{0} b_{1}^{0} $  & 3.0$	\times10^{-4}$\\ \hline 
$\bar{B}^{0} \to D^{0} h_{1} $  & 3.1$	\times10^{-4}$\\ \hline 
\end{tabular}

  \end{table}

  \clearpage

 \begin{table}

\centering
\caption{ Branching ratios for $B\to PA$ decays in CKM-favored mode involving $b\to c$ transition. The numbers in ( ) are for $\theta _{1} =-58^{\circ } $ and in [ ] are experimental values.}
 \label{t12}

\begin{tabular}{|l|l|} \hline 
\textbf{Decays} & \textbf{Branching ratios} \\ \hline 
\multicolumn{2}{|l|}{$\Delta b=1,  \Delta C=0,\Delta S=-1$} \\ \hline 
\multicolumn{1}{|l|}{$B^{-} \to K^{-} \chi _{c1} $} & 1.3$\times10^{-4}[(4.79\pm 0.23)\times 10^{-4} ]$ \\ \hline 
\multicolumn{1}{|l|}{$B^{-} \to D^{0} D_{s1}^{-} $} & 2.0$	\times10^{-3}$ \\ \hline 
\multicolumn{1}{|l|}{$B^{-} \to D^{0} \underline{D}_{s1}^{-} $} & 5.4$	\times10^{-4}$ \\ \hline 
\multicolumn{1}{|l|}{$B^{-} \to D_{s}^{-} D_{1}^{0} $} & 2.1$	\times10^{-3}$ \\ \hline 
\multicolumn{1}{|l|}{$B^{-} \to D_{s}^{-} \underline{D}_{1}^{0} $} & 3.9$	\times10^{-3}$ \\ \hline 
\multicolumn{1}{|l|}{$B^{-} \to \eta _{c} K_{1}^{-} $} & 2.5$	\times10^{-3}(2.5\times10^{-3})$ \\ \hline 
\multicolumn{1}{|l|}{$B^{-} \to \eta _{c} \underline{K}_{1}^{-} $} & 5.0$	\times10^{-5}$(7.0$\times10^{-5}$) \\ \hline 
\multicolumn{1}{|l|}{$\bar{B}^{0} \to \bar{K}^{0} \chi _{c1} $} & 1.2$\times10^{-4}$\newline $[(3.9\pm 0.4)\times 10^{-4} ]$ \\ \hline 
\multicolumn{1}{|l|}{$\bar{B}^{0} \to D^{+} D_{s1}^{-} $} & 1.9$	\times10^{-3}$ \\ \hline 
\multicolumn{1}{|l|}{$\bar{B}^{0} \to D^{+} \underline{D}_{s1}^{-} $} & 5.0$	\times10^{-4}$ \\ \hline 
\multicolumn{1}{|l|}{$\bar{B}^{0} \to D_{s}^{-} D_{1}^{+} $ } & 2.0$	\times10^{-3}$ \\ \hline 
\multicolumn{1}{|l|}{$\bar{B}^{0} \to D_{s}^{-} \underline{D}_{1}^{+} $ } & 3.7$	\times10^{-3}$ \\ \hline 
\multicolumn{1}{|l|}{$\bar{B}^{0} \to \eta _{c} \bar{K}_{1}^{0} $ } & 2.3$	\times10^{-3}$(2.3$\times10^{-3}$) \\ \hline 
\multicolumn{1}{|l|}{$\bar{B}^{0} \to \eta _{c} \underline{\bar{K}}_{1}^{0} $ } & 4.6$	\times10^{-5}$(6.5$\times10^{-5}$) \\ \hline 
\end{tabular}

 \end{table}

 \begin{table}
 
\centering
\caption{Branching ratios for $B\to PA$ decays in CKM-suppressed mode involving $b\to c$ transition. The values in the parenthesis are for $\theta _{1} =-58^{\circ } $.}
\label{t13}

\begin{tabular}{|l|l|} \hline 
\textbf{Decays} & \textbf{Branching ratios } \\ \hline 
\multicolumn{2}{|l|}{$\Delta b=1,  \Delta C=1,\Delta S=-1$ } \\ \hline 
$B^{-} \to K^{-} D_{1}^{0} $ & 6.4$	\times10^{-5}$\\ \hline 
$B^{-} \to K^{-} \underline{D}_{1}^{0} $  & 1.1$	\times10^{-4}$ \\ \hline 
$B^{-} \to D^{0} K_{1}^{-} $ & 1.8$	\times10^{-4}$(1.6$\times10^{-4}$) \\ \hline 
$B^{-} \to D^{0} \underline{K}_{1}^{-} $ & 6.3$	\times10^{-4}$(1.7$\times10^{-4}$) \\ \hline 
$\bar{B}^{0} \to \bar{K}^{0} D_{1}^{0} $ & 4.8$	\times10^{-8}$\\ \hline 
$\bar{B}^{0} \to \bar{K}^{0} \underline{D}_{1}^{0} $ & 6.8$	\times10^{-6}$\\ \hline 
$\bar{B}^{0} \to K^{-} D_{1}^{+} $ & 6.3$	\times10^{-5}$ \\ \hline 
$\bar{B}^{0} \to K^{-} \underline{D}_{1}^{+} $ & 1.6$	\times10^{-4}$ \\ \hline 
$\bar{B}^{0} \to D^{+} K_{1}^{-} $ & 4.7$	\times10^{-4}$(4.5$\times10^{-4}$) \\ \hline 
$\bar{B}^{0} \to D^{+} \underline{K}_{1}^{-} $ & 6.8$	\times10^{-4}$(1.3$\times10^{-4}$) \\ \hline 
$\bar{B}^{0} \to D^{0} \bar{K}_{1}^{0} $ & 7.7$	\times10^{-5}$(8.1$\times10^{-5})$ \\ \hline 
$\bar{B}^{0} \to D^{0} \bar{\underline{K}}_{1}^{0} $ & 4.1$	\times10^{-6}$(7.3$\times10^{-7})$ \\ \hline 
\end{tabular}

\end{table}

\begin{table}

\centering
\caption{Branching ratios for $B\to PA$ decays in CKM-suppressed mode involving $b\to c$ transition. Numbers in [ ] are experimental values.}
\label{t14}
\begin{tabular}{|l|l|} \hline 
\textbf{Decays} & \textbf{Branching ratios } \\ \hline 
\multicolumn{2}{|l|}{$\Delta b=1,  \Delta C=0,\Delta S=0$ } \\ \hline 
$B^{-} \to \pi ^{-} \chi _{c1} $ & 6.2$\times10^{-6}$\newline $[(2.2\pm 0.5)\times 10^{-5} ]$ \\ \hline 
$B^{-} \to D^{0} D_{1}^{-} $ & 1.2$	\times10^{-6}$ \\ \hline 
$B^{-} \to D^{0} \underline{D}_{1}^{-} $ & 1.6$	\times10^{-4}$ \\ \hline 
$B^{-} \to D^{-} D_{1}^{0} $ & 7.0$	\times10^{-5}$ \\ \hline 
$B^{-} \to D^{-} \underline{D}_{1}^{0} $ & 1.4$	\times10^{-4}$ \\ \hline 
$B^{-} \to \eta _{c} a_{1}^{-} $ &  9.4$	\times10^{-5}$\\ \hline 
$B^{-} \to \eta _{c} b_{1}^{-} $ &  7.0$	\times10^{-5}$\\ \hline 
$\bar{B}^{0} \to \pi ^{0} \chi _{c1} $ & 2.9$\times10^{-6}$\newline $[(1.12\pm 0.28)\times 10^{-5} ]$ \\ \hline 
$\bar{B}^{0} \to \eta \chi _{c1} $ &  1.4$	\times10^{-6}$\\ \hline 
$\bar{B}^{0} \to \eta '\chi _{c1} $  & 5.2$	\times10^{-7}$\\ \hline 
$\bar{B}^{0} \to D^{+} D_{1}^{-} $ & 1.1$	\times10^{-6}$ \\ \hline 
$\bar{B}^{0} \to D^{+} \underline{D}_{1}^{-} $ & 1.5$	\times10^{-4}$ \\ \hline 
$\bar{B}^{0} \to D^{-} D_{1}^{+} $ & 6.6$	\times10^{-5}$ \\ \hline 
$\bar{B}^{0} \to D^{-} \underline{D}_{1}^{+} $ & 1.3$	\times10^{-4}$ \\ \hline 
$\bar{B}^{0} \to \eta _{c} a_{1}^{0} $  &  4.4$	\times10^{-5}$\\ \hline 
$\bar{B}^{0} \to \eta _{c} f_{1} $  &  3.9$	\times10^{-5}$\\ \hline 
$\bar{B}^{0} \to \eta _{c} b_{1}^{0} $ &  3.2$	\times10^{-5}$\\ \hline 
$\bar{B}^{0} \to \eta _{c} h_{1} $ &  3.5$	\times10^{-5}$\\ \hline 
\end{tabular}

\end{table}
\clearpage

\begin{table}

\centering
\caption{Branching ratios for $B\to PA$ decays involving $b\to u$ transition. The numbers in ( ) are for $\theta _{1} =-58^{\circ } $ and in [ ] are experimental values.}
\label{t15}
\begin{tabular}{|l|l|} \hline 
\textbf{Decays} & \textbf{Branching ratios} \\ \hline 
\multicolumn{2}{|l|}{$\Delta b=1,  \Delta C=-1,\Delta S=-1$ } \\ \hline 
$B^{-} \to \pi ^{0} D_{s1}^{-} $ & 7.7$	\times10^{-6}$\\ \hline 
$B^{-} \to \pi ^{0} \underline{D}_{s1}^{-} $  & 2.7$	\times10^{-6}$\\ \hline 
$B^{-} \to \eta D_{s1}^{-} $ & 3.9$	\times10^{-6}$\\ \hline 
$B^{-} \to \eta \underline{D}_{s1}^{-} $ & 1.4$	\times10^{-6}$\\ \hline 
$B^{-} \to K^{-} \bar{D}_{1}^{0} $ & 2.2$	\times10^{-9}$\\ \hline 
$B^{-} \to K^{-} \underline{\bar{D}}_{1}^{0} $ & 1.4$	\times10^{-6}$\\ \hline 
$B^{-} \to \eta 'D_{s1}^{-} $ & 1.9$	\times10^{-6}$\\ \hline 
$B^{-} \to \eta '\underline{D}_{s1}^{-} $ & 6.6$	\times10^{-7}$\\ \hline 
$B^{-} \to \bar{D}^{0} K_{1}^{-} $ & 1.2$\times10^{-5}$(1.3$\times10^{-5})$ \\ \hline 
$B^{-} \to \bar{D}^{0} \underline{K}_{1}^{-} $ & 6.7$\times10^{-7}$(1.2$\times10^{-7})$ \\ \hline 
$B^{-} \to D_{s}^{-} a_{1}^{0} $ & 1.4$\times10^{-4}$\newline $[<1.8\times 10^{-3} ]$ \\ \hline 
$B^{-} \to D_{s}^{-} f_{1} $ & 1.3$	\times10^{-4}$\\ \hline 
$B^{-} \to D_{s}^{-} b_{1}^{0} $ & 7.7$	\times10^{-5}$\\ \hline 
$B^{-} \to D_{s}^{-} h_{1} $ &  8.1$	\times10^{-5}$\\ \hline 
$\bar{B}^{0} \to \pi ^{+} D_{s1}^{-} $  & 1.4$	\times10^{-5}$\\ \hline 
$\bar{B}^{0} \to \pi ^{+} \underline{D}_{s1}^{-} $  & 5.1$	\times10^{-6}$\\ \hline 
$\bar{B}^{0} \to \bar{K}^{0} \bar{D}_{1}^{0} $  & 2.1$	\times10^{-9}$\\ \hline 
$\bar{B}^{0} \to \bar{K}^{0} \underline{\bar{D}}_{1}^{0} $  & 1.3$	\times10^{-6}$\\ \hline 
$\bar{B}^{0} \to \bar{D}^{0} \bar{K}_{1}^{0} $ & 1.1$\times10^{-5}$(1.2$\times10^{-5})$ \\ \hline 
$\bar{B}^{0} \to \bar{D}^{0} \underline{\bar{K}}_{1}^{0} $ & 6.3$\times10^{-7}$(1.1$\times10^{-7})$ \\ \hline 
$\bar{B}^{0} \to D_{s}^{-} a_{1}^{+} $ & 2.7$\times10^{-4}$\newline $[<2.1\times 10^{-3} ]$ \\ \hline 
$\bar{B}^{0} \to D_{s}^{-} b_{1}^{+} $ & 1.4$	\times10^{-4}$\\ \hline 
\end{tabular}
\end{table}
\clearpage
\clearpage

\begin{table}

\centering
\caption{Branching ratios for $B\to PA$ decays involving $b\to u$  transition}\label{t16}
\begin{tabular}{|c|c|} \hline 
\textbf{Decays} & \textbf{Branching ratios} \\ \hline 
\multicolumn{2}{|l|}{$\Delta b=1,  \Delta C=-1,\Delta S=0$} \\ \hline 
$B^{-} \to \pi ^{0} D_{1}^{-} $ & 1.2$	\times10^{-9}$\\ \hline 
$B^{-} \to \pi ^{0} \underline{D}_{1}^{-} $ & 6.1$	\times10^{-7}$\\ \hline 
$B^{-} \to \pi ^{-} \bar{D}_{1}^{0} $  & 1.3$	\times10^{-10}$\\ \hline 
$B^{-} \to \pi ^{-} \underline{\bar{D}}_{1}^{0} $  & 6.6$	\times10^{-8}$\\ \hline 
$B^{-} \to \eta D_{1}^{-} $ & 6.6$	\times10^{-10}$ \\ \hline 
$B^{-} \to \eta \underline{D}_{1}^{-} $  & 3.1$	\times10^{-7}$\\ \hline 
$B^{-} \to \eta 'D_{1}^{-} $ & 2.9$	\times10^{-10}$\\ \hline 
$B^{-} \to \eta '\underline{D}_{1}^{-} $ & 1.5$	\times10^{-7}$\\ \hline 
$B^{-} \to D^{-} a_{1}^{0} $  & 4.5$	\times10^{-6}$\\ \hline 
$B^{-} \to D^{-} f_{1} $  & 4.1$	\times10^{-6}$\\ \hline 
$B^{-} \to D^{-} b_{1}^{0} $  & 2.4$	\times10^{-6}$\\ \hline 
$B^{-} \to D^{-} h_{1} $ & 2.5$	\times10^{-6}$\\ \hline 
$B^{-} \to \bar{D}^{0} a_{1}^{-} $ & 5.0$	\times10^{-7}$\\ \hline 
$B^{-} \to \bar{D}^{0} b_{1}^{-} $  & 2.6$	\times10^{-7}$\\ \hline 
$\bar{B}^{0} \to \pi ^{+} D_{1}^{-} $ & 2.2$	\times10^{-9}$\\ \hline 
$\bar{B}^{0} \to \pi ^{+} \underline{D}_{1}^{-} $ & 1.1$	\times10^{-6}$\\ \hline 
$\bar{B}^{0} \to \pi ^{0} \bar{D}_{1}^{0} $ & 5.9$	\times10^{-11}$\\ \hline 
$\bar{B}^{0} \to \pi ^{0} \underline{\bar{D}}_{1}^{0} $ & 3.1$	\times10^{-8}$\\ \hline 
$\bar{B}^{0} \to \eta \bar{D}_{1}^{0} $ & 3.0$	\times10^{-11}$\\ \hline 
$\bar{B}^{0} \to \eta \underline{\bar{D}}_{1}^{0} $ & 1.6$	\times10^{-8}$\\ \hline 
$\bar{B}^{0} \to \eta '\bar{D}_{1}^{0} $ & 1.5$	\times10^{-11}$\\ \hline 
$\bar{B}^{0} \to \eta '\underline{\bar{D}}_{1}^{0} $ & 7.6$	\times10^{-9}$\\ \hline 
$\bar{B}^{0} \to D^{-} a_{1}^{+} $  & 8.4$	\times10^{-6}$\\ \hline 
$\bar{B}^{0} \to D^{-} b_{1}^{+} $  & 4.4$	\times10^{-6}$\\ \hline 
$\bar{B}^{0} \to \bar{D}^{0} a_{1}^{0} $ & 2.3$	\times10^{-7}$\\ \hline 
$\bar{B}^{0} \to \bar{D}^{0} f_{1} $ & 2.1$	\times10^{-7}$\\ \hline 
$\bar{B}^{0} \to \bar{D}^{0} b_{1}^{0} $ & 1.2$	\times10^{-7}$\\ \hline 
$\bar{B}^{0} \to \bar{D}^{0} h_{1} $ & 1.3$	\times10^{-7}$\\ \hline 
\end{tabular}

\end{table}
\clearpage

\begin{table}

\centering
\caption{HQS constrained branching ratios for $B\to PA$ decays in CKM-favored mode involving $b\to c$ transition. Numbers in [ ] are experimental values.}\label{t17}
\begin{tabular}{|c|l|c|} \hline 
\multirow{2}{*}{\textbf{Decays}} & \multicolumn{2}{c|}{\textbf{Branching ratios}} \\ \hhline{~--}
  & \textbf{Our Results} & \textbf{\cite{18,19}} \\ \hline 
\multicolumn{2}{|l|}{$\Delta b=1,  \Delta C=1,\Delta S=0$ } &  \\ \hline 
$B^{-} \to \pi ^{-} D_{1}^{0} $  & 2.7$	\times10^{-3}$[$(7.5\pm 1.7)\times 10^{-4} $] & 3.7$	\times10^{-4}$\\ \hline 
$B^{-} \to \pi ^{-} \underline{D}_{1}^{0} $  & 1.4$	\times10^{-3}$[$(1.5\pm 0.6)\times 10^{-3} $] & 1.1$	\times10^{-3}$\\ \hline 
$\bar{B}^{0} \to \pi ^{0} D_{1}^{0} $  & 2.0$	\times10^{-5}$ &  \\ \hline 
$\bar{B}^{0} \to \pi ^{0} \underline{D}_{1}^{0} $  & 9.3$	\times10^{-5}$ &  \\ \hline 
$\bar{B}^{0} \to \pi ^{-} D_{1}^{+} $  & 3.2$	\times10^{-3}$ & 6.8$	\times10^{-4}$\\ \hline 
$\bar{B}^{0} \to \pi ^{-} \underline{D}_{1}^{+} $  & 5.0$	\times10^{-4}$ & 1.0$	\times10^{-3}$\\ \hline 
$\bar{B}^{0} \to \eta D_{1}^{0} $  & 1.0$	\times10^{-5}$ &  \\ \hline 
$\bar{B}^{0} \to \eta \underline{D}_{1}^{0} $  & 4.8$	\times10^{-5}$ &  \\ \hline 
$\bar{B}^{0} \to \eta 'D_{1}^{0} $  & 4.8$	\times10^{-6}$ &  \\ \hline 
$\bar{B}^{0} \to \eta '\underline{D}_{1}^{0} $  & 2.3$	\times10^{-5}$ &  \\ \hline 
\multicolumn{2}{|l|}{$\Delta b=1,  \Delta C=0,\Delta S=-1$} &  \\ \hline 
$B^{-} \to D_{s}^{-} D_{1}^{0} $$ $ & 8.2$	\times10^{-3}$ & 9.6$	\times10^{-4}$\\ \hline 
$B^{-} \to D_{s}^{-} \underline{D}_{1}^{0} $$ $ & 1.2$	\times10^{-3}$  & 1.3$	\times10^{-3}$\\ \hline 
$B^{-} \to D^{0} D_{s1}^{-} $ & 2.8$	\times10^{-3}$& 4.3$	\times10^{-3}$\\ \hline 
$B^{-} \to D^{0} \underline{D}_{s1}^{-} $ & 8.5$	\times10^{-4}$& 3.1$	\times10^{-4}$\\ \hline 
$\bar{B}^{0} \to D_{s}^{-} D_{1}^{+} $  & 7.6$	\times10^{-3}$ & 8.8$	\times10^{-4}$\\ \hline 
$\bar{B}^{0} \to D_{s}^{-} \underline{D}_{1}^{+} $  & 1.2$	\times10^{-3}$ & 1.2$	\times10^{-3}$\\ \hline 
$\bar{B}^{0} \to D^{+} D_{s1}^{-} $ & 2.6$	\times10^{-3}$& 3.9$	\times10^{-3}$ \\ \hline 
$\bar{B}^{0} \to D^{+} \underline{D}_{s1}^{-} $ & 7.9$	\times10^{-4}$& 2.8$	\times10^{-4}$\\ \hline 
\end{tabular}
\end{table}
\clearpage

\begin{table} 

\centering
\caption{HQS constrained branching ratios for $B\to PA$ decays in CKM-suppressed mode involving $b\to c$ transition}\label{t18}
\begin{tabular}{|c|c|} \hline 
\textbf{Decays} & \textbf{Branching ratios } \\ \hline 
\multicolumn{2}{|l|}{$\Delta b=1,  \Delta C=1,\Delta S=-1$ } \\ \hline 
$B^{-} \to K^{-} D_{1}^{0} $ & 2.1$	\times10^{-4}$\\ \hline 
$B^{-} \to K^{-} \underline{D}_{1}^{0} $  & 9.5$	\times10^{-5}$ \\ \hline 
$\bar{B}^{0} \to \bar{K}^{0} D_{1}^{0} $ & 2.3$	\times10^{-6}$\\ \hline 
$\bar{B}^{0} \to \bar{K}^{0} \underline{D}_{1}^{0} $ & 1.1$	\times10^{-5}$\\ \hline 
$\bar{B}^{0} \to K^{-} D_{1}^{+} $ & 2.4$	\times10^{-4}$ \\ \hline 
$\bar{B}^{0} \to K^{-} \underline{D}_{1}^{+} $ & 3.7$	\times10^{-5}$ \\ \hline 
\multicolumn{2}{|l|}{$\Delta b=1,  \Delta C=0,\Delta S=0$} \\ \hline 
$B^{-} \to D^{0} D_{1}^{-} $ & 5.3$	\times10^{-5}$ \\ \hline 
$B^{-} \to D^{0} \underline{D}_{1}^{-} $ & 2.5$	\times10^{-4}$ \\ \hline 
$B^{-} \to D^{-} D_{1}^{0} $  & 2.7	$\times10^{-4}$ \\ \hline 
$B^{-} \to D^{-} \underline{D}_{1}^{0} $ & 4.1$	\times10^{-5}$ \\ \hline 
$\bar{B}^{0} \to D^{+} D_{1}^{-} $ & 4.9$	\times10^{-5}$ \\ \hline 
$\bar{B}^{0} \to D^{+} \underline{D}_{1}^{-} $ & 2.3$	\times10^{-4}$ \\ \hline 
$\bar{B}^{0} \to D^{-} D_{1}^{+} $ & 2.5$	\times10^{-4}$ \\ \hline 
$\bar{B}^{0} \to D^{-} \underline{D}_{1}^{+} $ & 3.8$	\times10^{-5}$ \\ \hline 
\end{tabular}
\end{table}


\begin{thebibliography}{99}
\bibitem{38}  Experimental averages are given in Table V of ref.[16].
\end{thebibliography}
\end{document}